\documentclass[manuscript]{aastex}
\usepackage{graphicx}
\usepackage{natbib}
\bibpunct{(}{)}{,}{a}{}{;}

\newcommand{\vlsr}{$V_{\mbox{\scriptsize LSR}}$}
\newcommand{\kms}{\,km\,s$^{-1}$}
	
\newcommand{\etal}{~et~al.}
\newcommand{\water}{H$_{2}$O}
\newcommand{\target}{I18286}

\slugcomment{To be submitted to ApJ}

\shorttitle{Precessing jet from IRAS\,18286$-$0959}
\shortauthors{Yung et al.}

\begin{document}


\title{High Velocity Precessing Jets from the Water Fountain 
       IRAS\,18286$-$0959 Revealed by VLBA Observations}


\author{Bosco H. K. Yung\altaffilmark{1}, Jun-ichi Nakashima\altaffilmark{1},
        Hiroshi Imai\altaffilmark{2}, Shuji Deguchi\altaffilmark{3}, \\
        Philip J. Diamond\altaffilmark{4}$^{,}$\altaffilmark{5}
        and Sun Kwok\altaffilmark{1}}





\altaffiltext{1}{Department of Physics, The University of Hong Kong,
                 Pokfulam Rd, Hong Kong, China}
\altaffiltext{2}{Graduate School of Science and Engineering, Kagoshima 
                 University, Kagoshima 890-0065, Japan}
\altaffiltext{3}{Nobeyama Radio Observatory, National Astronomical Observatory
                 of Japan, Minamimaki, Minamisaku, Nagano 384-1305, Japan}
\altaffiltext{4}{Jodrell Bank Centre for Astrophysics, Alan Turing Building,
                 University of Manchester, Manchester M13 9PL, United Kingdom}
\altaffiltext{5}{CSIRO Astronomy and Space Science, PO Box 76, Epping, 
                 NSW 1710, Australia}


\begin{abstract}

We report the results of multi-epoch VLBA observations of the 22.2\,GHz 
\water\ maser emission associated with the ``water fountain''  
IRAS\,18286$-$0959.
We suggest that this object is the second example of 
a highly collimated bipolar \textsl{precessing} outflow traced by \water\
maser emission, the other is 
W\,43A. The detected \water\ emission peaks 
are distributed over a velocity range from $-50$\kms\ to 150\kms.  
The spatial distribution of over 70\% of the identified maser features 
is found to be highly 
collimated along a spiral jet~(jet~1) extended southeast to northwest, 
the remaining features appear to trace another 
spiral jet~(jet~2) with a different orientation. The two jets form 
a ``double-helix'' pattern which lies across $\sim$200 milliarcseconds. 
The maser distribution is reasonably fit by a model consisting of two bipolar
precessing jets.
The 3D velocities of jet~1 and jet~2 are derived to be 138\kms\ and 99\kms,
respectively. The precession period of jet~1 is about 56 years. For jet~2, 
three possible models are tested and they
give different values for the kinematic parameters. We propose that 
the appearance of two jets 
is the result of a single driving source with significant proper motion.

\end{abstract}


\keywords{masers --- stars: AGB and post-AGB --- stars: winds, outflows
          --- stars: evolution}



\section{Introduction}
\label{sec:intro}

An asymptotic giant branch~(AGB) star is usually spherical in shape, while a 
planetary nebula often exhibits a bipolar or even multi-polar morphology 
\citep[][and references therein]{kwok08iaup,kwok10pasa}.  
The mechanism that causes
the morphological change is unclear, however it is suggested 
that high velocity outflows originating from AGB or post-AGB stars are closely 
related to the shaping of planetary nebulae \citep{sahai98aj}. 

The term ``water fountain'' is used to describe a fast, collimated, 
molecular outflow that can be traced by \water\ maser emission
\citep{likkel88apj}. The outflow velocity is larger 
than the typical expansion velocity of an OH/IR 
star's circumstellar envelope ($10-20$\kms) revealed by 
1612\,MHz OH maser emission often associated with evolved stars 
\citep{hekkert89aas}. From the view point of stellar evolution, this type of 
object is in the transition phase from an AGB star to the central star of a 
planetary nebula, during this phase the mass loss rate reaches its maximum.
It is believed that water fountains are closely related to the shaping of 
planetary nebulae \citep{imai07iaup}, and the high mass-loss rate during this 
stage contributes significantly to the redistribution of matter from the star
to interstellar space. IRAS\,16342$-$3814 \citep{sahai99apj,claussen09apj}, 
W\,43A \citep{diamond88iaup,imai02nature}, and IRAS\,19134$+$2131 
\citep{imai04aa,imai07apj} are representative examples of water fountain 
sources. IRAS\,19190$+$1102 \citep{day10apj} and IRAS\,18113$-$2503 
\citep{gomez11apjl} are arguably the most recent members of this class of 
object.  

There are other discoveries with regard to water fountains
besides \water\ and OH masers. The first detection of SiO maser 
emission associated with W\,43A was reported by 
\citet{nakashima03pasj}, suggesting 
there has been on-going, copious mass loss from the stellar 
surface. \citet{imai05apj} show that the distribution of
the SiO maser features are well fitted by a biconically expanding model, 
which is not typical in AGB stars. \citet{he08aa} observed the 
CO~$J=2-1$ emission in IRAS\,16342$-$3814 and \citet{imai09pasj}   
found a very high velocity ($\sim$200\kms) CO~flow
in CO~$J=3-2$ emission from the same object.
These results suggest that the high velocity outflow plays a significant role 
in the stellar mass loss.  

The jets from water fountains are often highly collimated. 
\citet{vlemmings06nature} and \citet{amiri10aa} measured the magnetic field
around W\,43A and suggested that it plays an important role in
producing the collimated jet. To date this is the only water 
fountain where a magnetic field has been detected associated with the jet. 
W\,43A is also the first water fountain in which a precessing jet is 
observed \citep{imai02nature,imai05apj}. 
A similar jet pattern is found in an optical image of IRAS\,16342$-$3814 
\citep{sahai05apj}, with the authors proposing that it is a ``corkscrew jet'', 
in which the 
ejecta move along a spiral path rather than having a ballistic motion as in 
the precessing case. However, the detailed morpho-kinematical structure of the 
\water\ masers in IRAS\,16342$-$3814 is still unknown due to the 
limited number of detected \water\ maser features.
Note that we define here a ballistic 
motion as a linear, constant velocity motion.

Other than the examples mentioned above, there is another
distinctive water fountain candidate: 
IRAS\,18286$-$0959 (hereafter abbreviated as \target).
It is one of the OH maser sources observed by \citet{sevenster97aas}, and 
then by \citet{imai08evn}.
The \water\ maser emission of \target\ was first detected 
with the NRO~45\,m telescope \citep{deguchi07apj}. Its single-dish 
spectrum shows many emission peaks throughout the velocity range 
$-50$\kms\ to $150$\kms\ with respect to the local standard of
rest~(LSR). The spectrum resembles the spectral profile of \water\ maser 
emission towards a young stellar object~(YSO),
nonetheless, \target\ is expected to be an AGB or a post-AGB star
(see Section~\ref{ssec:evo}).
Therefore, in terms of both the jet kinematics 
and the evolutionary status, this object has the aforementioned properties of 
a water fountain. The first VLBA observation of this object was undertaken  
shortly after its discovery \citep{imai07iaup}. 
Over 100 maser features were identified in a region extending more 
than a hundred milliarcseconds~(mas). 

In this paper, new VLBA results on \target\ are presented.
The aim of this project is to investigate the kinematics of the high velocity 
component of \target\ through the motions of \water\ maser features, and to
discuss the evolutionary status of this object.
Details of the observations and data reduction are described 
in Section~\ref{sec:obs}. The observational results are presented in 
Section~\ref{sec:res}. The kinematic models for the 
jet are given in Section~\ref{sec:model}, followed by the discussion in 
Section~\ref{sec:dis}. 

We would like to clarify
the difference between ``maser spots'' and `` maser features'', as these terms
will appear many times in this paper.
A maser feature is composed of a physical clump of gas
that emits a maser. The received flux usually spans a small range of
frequency due to the effect of Doppler broadening. Hence the emission from a
maser feature will be spread across a few (usually less than 10, depending 
on the frequency resolution in use) velocity channels. In contrast, a maser 
spot is the emission from one
of the velocity channels in such a maser feature. In other words, a maser 
feature consists of several 
maser spots; a maser spot represents the flux of one velocity component of the 
feature received in a single velocity channel. However, in order to simplify
the analysis procedure, we will represent a maser feature by its brightest
maser spot (see Section~\ref{sec:res} for the details).

\section{Observations and Data Reduction}
\label{sec:obs}

Table~\ref{tab:status} gives a summary of the status of the VLBA observations 
of \target\ \water\ masers and the data reduction. The observations were 
made at 6~epochs (epochs~A to F) over a year from 2008~April~21 to 
2009~May~19. 
The duration of each observation 
was 6~hours in total, including scans on the calibrator OT\,081, which 
was observed for 4~minutes in every 40~minutes for calibration of clock 
offsets and bandpass characteristics. For high accuracy astrometry, a
phase-referencing mode was adopted, in which each 
antenna nodded between the phase-reference source ICRF~J183220.8$-$103511 
and the target maser source in a cycle of 60\,s. Scans in the geodetic VLBI 
observation mode were also included for 30~minutes at the beginning and end 
of the observation. The resulting integration time for \target\ was about 
80~minutes. In this paper, only the results obtained on the basis 
of a self-calibration procedure are presented. 
 
The received signals were recorded at a rate of 128\,Mbits\,s$^{-1}$ with 
2~bits per sample into two base-band channels (BBCs) for dual circular 
polarization signals. The total BBC band width was set to 16\,MHz, 
corresponding to a BBC \vlsr\ range of 
215.9\kms. The center velocity of the BBCs was set to 50\kms. 
The recorded data were correlated with 
the VLBA correlator at Socorro using an accumulation period of 2\,s. The 
data of each BBC were divided into 1024 spectral channels, yielding a 
velocity spacing of 0.21\kms\ per spectral channel. Based on a separate 
astrometric observation (Imai\etal, in preparation), the 
following coordinates of \target\ were adopted as the delay-tracking center 
in the data correlation:
$\alpha_{\rm J2000}=$18$^{h}$31$^{m}$22$^{s}$\hspace{-2pt}.934,
$\delta_{\rm J2000}=$~$-$09$^{\circ}$57$^{\prime}$21\farcs 70.

For the VLBA data analysis, the NRAO's AIPS package and MIRIAD 
\citep{sault95asp} were used for visibility data calibration and image cube 
synthesis, respectively. For the astrometry, the phase-referencing technique
was applied \citep[e.g.,][]{beasley95asp}. 
However, in order to obtain higher quality maser image cubes that are not 
affected by deformation of images due to imperfect phase-referencing, the 
visibility data were self-calibrated using a spectral channel containing bright
maser emission as the reference. Column~3 of Table~\ref{tab:status} lists 
the LSR velocity of the spectral channel for the self-calibration. Columns~4 
and 5 in Table~\ref{tab:status} list the rms noise level in the maser 
cubes (in spectral channels without bright maser emission) and the 
synthesized beam parameters. It is noted that most of the maser spots could 
be detected with only baselines shorter than 3000\,km, leading to a larger 
effective synthesized beam size. Thanks to the proximity of the 
phase-referenced maser spot to the delay-tracking center ($<300$\,mas), 
position drifts of maser spots attributed to the phase drift 
with frequency due to group-delay residuals, were negligible.

\section{Results}
\label{sec:res}

\subsection{Spectra}

Figure~\ref{fig:afspectra} shows the VLBA spectra of \target\
\water\ maser emission in epochs~A and F. The spectra were created by 
integrating the flux for all the velocity channels in the regions containing
maser emission, using the \textit{imspec} task in MIRIAD.  
Unlike W\,43A, IRAS\,19134$+$2131, IRAS\,16342$-$3814 \citep{likkel92apj} 
and IRAS\,16552$-$3050 \citep{suarez08apj} in which the \water\ maser 
spectra clearly show two clusters of emission peaks widely separated in 
velocity space (hereafter referred to as a ``double-group profile'' for convenience, 
note that they may not have exactly two peaks), the spectra of 
\target\ show emission peaks throughout the velocity range
from $-50$\kms\ to 150\kms. 
This profile characteristic is similar to that of
OH\,009.1$-$0.4 \citep{walsh09mnras}, which is another known water fountain
with many spectral groups of peaks. 
Figure~\ref{fig:nrospectra} shows the spectra obtained from the NRO~45\,m 
telescope observations in 2006, 2008 and 2010.
It can be seen that the spectral profile changes rapidly with time but the 
overall velocity range of the \water\ maser emission remains the same over the 
four years \citep[cf.][]{imai07iaup}. However, 
it is possible that there exist a few weak 
components with velocity greater than 150\kms, which have not been detected in
the VLBA observations. 
The integrated flux values from the single-dish and VLBA spectra are within 
the usual range of such radio observations ($\sim$30\%). It allows us to be 
confident that the missing flux in our interferometric observations is not 
significant.



\subsection{Spatial Distribution}
\label{ssec:spatial}

The number of identified maser features in the VLBA observations 
was different from epoch to epoch, as listed in Table~\ref{tab:spots}. 
The maximum number of features found in a single epoch was 143,
which is more than the number collected from the first VLBA observation 
\citep{imai07iaup}. Few maser features survived over all the 6 epochs, but 
many existed in 2 to 3 consecutive epochs. 
In the current analysis, it is found that
each of the features spans a velocity
range of less than 10 spectral channels. The spatial position is obtained 
by performing a 2-dimensional Gaussian fit at the channel where the 
maximum flux of each feature is found (i.e. the brightest maser spot). 
For simplicity, this brightest maser spot of each feature is used in the 
proper motion calculation, and it is assumed representative of its 
corresponding feature. 
Therefore from now on whenever the term ``maser feature'' is used, it 
means ``the brightest maser spot of the feature''. The maser spots 
have relatively simple brightness structures so most of them have similar 
sizes to that of the synthesized beam. Figure~\ref{fig:maserC} shows the spatial 
distribution of the maser 
emission peaks. They extend across a region of about 200\,mas, with the 
red-shifted and blue-shifted clusters of the peaks located at the 
southeast and northwest sides respectively. \target\ lies at a distance of 
approximately 4.0\,kpc according to an annual parallax measurement
(Imai\etal, in preparation), so 200\,mas is equivalent to a linear scale of 
$\sim$800\,AU.

Some characteristics of the outflow pattern are shown in 
Figure~\ref{fig:maserC}, they are summarized as follows:
\begin{enumerate} 
\item Most of the maser features (over 80\% of the total number) are 
      concentrated in two arcs stretching from the center of the 
      structure to both the red-shifted and blue-shifted ends. 
      The two arcs are located almost point-symmetrically with respect to the 
      central part of the distribution.
      The line-of-sight velocities (\vlsr) change approximately continuously 
      along the arcs. The features near to the center of the structure
      have \vlsr\ close to the systemic velocity ($\sim50$\kms), while those 
      further away have \vlsr\ with larger offsets 
      from the systemic velocity.
      This characteristic is unique to \target\ amongst all the known water
      fountains.  
\item Some features, though having similar spatial distances from the center,
      are shown to have significantly different \vlsr\ (difference $>50$\kms). 
      They are observed to be overlapping on both the northern and southern sides of 
      the structure (Region~I and II). 
\item There are also a few maser features found in the areas loosely 
      defined by Void~I and II, which are separated from the two arcs.
\end{enumerate}
Finally, there was one maser feature that survived in all epochs.
Its \vlsr\ in epochs A to F is 52.01\kms, 51.58\kms, 51.57\kms, 51.57\kms,
51.58\kms\ and 51.58\kms, respectively. These values are in the
middle of the whole velocity range of the emission spectrum. In addition, 
this feature is also near to the spatial center of the 
structure. Therefore its brightest maser spot has been 
chosen as a reference point (indicated as the origin in Figure 
\ref{fig:maserC}); this enables easy comparison between different epochs 
when finding the proper motion of maser features.


\subsection{Proper Motions}

Figure \ref{fig:proper} shows the proper motions of 54 \water\ maser 
features identified in \target. Most of them did not survive more than 
3 epochs of the observations, 
with the exception of the chosen reference feature. 
The proper motions are therefore determined by 
tracing the features that existed in any 3 consecutive epochs, the 
velocities (as illustrated by the length of the vectors) are 
calculated by measuring the shift in position divided by the time over the 
mentioned period. The coordinates of the 54 maser features (represented by 
the brightest maser spot in each of them) in 3 different 
epochs are listed in Table~\ref{tab:proper}. 
The proper motions reveal the 
bipolar nature of this water fountain, which is similar to W\,43A 
\citep{imai02nature} and IRAS\,19134$+$2131 \citep{imai07apj}. 
The vectors all originate
from the central region of the structure but it is obvious that they do
not converge to a single point and there are notable deviations on the 
vectors' directions (see Section~\ref{ssec:kstest}). 
Table~\ref{tab:prop_vel} lists
the velocity information of these maser features; they are divided into two
groups, which correspond to the proposed jet~1 and jet~2, respectively.
More details are given in later sections.

\section{Kinematic Modeling}
\label{sec:model}

It is difficult to propose a unique model using only the data available 
at the current time. However, any plausible model should be able to reproduce 
the three main characteristics of \target\ listed in 
Section~\ref{ssec:spatial}. In the following, we examine models using these 
characteristics.

\subsection{The Bipolar Precessing Jet Model}
\label{ssec:pre}

We propose a bipolar precessing jet model based on the fact that it can 
reproduce all three aforementioned characteristics.
The bipolar jet is assumed to be point-symmetrical in the central part of 
the maser feature distribution.
Qualitatively, precession will generate a helical jet pattern 
\citep[e.g. the model of W\,43A in][]{imai02nature}; the arc-shaped 
distribution of maser features can be explained by this phenomenon. 
In addition, as the jet precesses, the gas molecules are ejected in a 3D
direction that varies at different instants. Therefore different regions of the
arcs do not have the same \vlsr, which is just the projected velocity along 
the line-of-sight. The first characteristic is therefore well explained. 

We propose that that the second and third characteristics could be reproduced 
by two modifications to the simple precessing jet model: (1) instead of a 
stationary driving source, we assume one that is moving; (2) we assume the ejection is 
episodic. These are reasonable assumptions as a source with significant proper
motion is common, and episodic ejections in evolved stellar objects such as 
planetary nebulae have been observed before 
\citep[e.g.][]{lopez95apj}. Alternatively, we could assume the existence of two driving 
stars, rather than a single source with secular motion. Both models are explored in the 
discussion that follows.

The maser features in region~I of 
Figure~\ref{fig:maserC} can be divided into two groups based on \vlsr: those 
represented in ``grey'', with \vlsr\ roughly around 20\kms\ to 30\kms; and 
those represented in ``blue'', with \vlsr\,$<0$. In region~II, the same 
division is possible, in which there are distinct ``yellow'' and ``red'' maser 
features. The colorbar clearly shows that the two groups in both regions
have very different \vlsr. We suggest the two groups are formed by different 
ejections of an episodic outflow, consistent
with the above assumptions. For simplicity, we fit the jet pattern in the two 
ejection episodes independently by using two apparent jets.

The basic set of parametric equations for one bipolar jet is:
\begin{eqnarray}
X_{\mathrm{R.A.}} &=& 
           -V_{\mathrm{jet}}t\cos(\frac{2\pi t}{T})\sin\alpha+X_{0}~;
           \label{eq:x}\\
Y_{\mathrm{Dec.}} &=& 
            V_{\mathrm{jet}}t\cos\alpha+Y_{0}~;\label{eq:y}\\
Z_{\mathrm{LOS}} &=& 
           -V_{\mathrm{jet}}|t|\sin(\frac{2\pi t}{T})\sin\alpha+Z_{0}~,
           \label{eq:z}
\end{eqnarray}
where $X_{\mathrm{R.A.}}$, $Y_{\mathrm{Dec.}}$ 
and $Z_{\mathrm{LOS}}$ (LOS stands for line-of-sight from the local standard
of rest) are spatial coordinates of any
point on the jet in 3D space, and $t$ is the time of travel to the 
supposed point. $t$ also acts as a running parameter linking
eq.~(\ref{eq:x}) to (\ref{eq:z}): when $t$ is positive, the equations 
represent the
northern part of the bipolar jet; when $t$ is negative, the southern part is
being considered. This set of equations is derived from 
the equations of a helix. A helix is chosen because mathematically this
represents the spatial and kinematical pattern formed by a precession jet.
The projected (2D) curves from the two model jets are used to fit the spatial 
and \vlsr\ 
distribution of the \water\ maser features in each 
epoch independently. The free parameters for any model of the jets are the 
systemic LSR velocity of the jet ($V_{\mathrm{sys}}$, not shown in the
equations, see below), precession period~($T$), 
precession angle~($\alpha$), 
position angle~(P.A.) of the jet axis, the 
inclination angle~(I.A.) of the axis with respect to the sky plane, and 
the position of the driving source, $X_{0}$ and $Y_{0}$. 
Since we are fitting the 2D projection only, $Z_{0}$ is set to 0. 
The jet velocity, $V_{\mathrm{jet}}$, is determined in the following way: 
the maser features listed in Table~\ref{tab:prop_vel} are divided into two 
groups for each bipolar jet, one group is for the features that are 
travelling to the north, the other group is for those travelling to the 
south. The mean 3D velocity of each group is calculated, and 
$V_{\mathrm{jet}}$ is derived from the difference of the group velocities 
divided by 2.
$V_{\mathrm{jet}}$ is assumed to be constant over the observation period.
For jet~1, $V_{\mathrm{jet}}$ is equal to $138\pm41$\kms, and for 
jet~2 it is equal to $99\pm21$\kms.
The resultant helix pattern can 
be spatially transformed and rotated (by Euler's rotation) in order to find 
the position of the driving source and the jet's axis direction 
such that the 2D projection of the model is best fitted with the 
observed maser distribution. The P.A. and I.A. can then be deduced from the 
Euler angles. 
Let $X'_{\mathrm{R.A.}}$, $Y'_{\mathrm{Dec.}}$ and $Z'_{\mathrm{LOS}}$ be
the new coordinates of each point on the jet model after any spatial 
transformation. Then the \vlsr\ of each point is given by
\begin{equation}
   V_{\mathrm{LSR}}=\frac{Z'_{\mathrm{LOS}}}{t} + V_{\mathrm{sys}}~.
\label{eq:vlsr}
\end{equation}
For the fitting, the dimensionless coefficient of determination, 
$R^{2}$, is used to measure the goodness of fit, where 
$R^{2} \leq 1$ \citep{steel60}. 

In a normal curve fitting process, the data points only carry spatial 
information. However in our case, in addition to the R.A. and Dec., each point 
is also associated with a \vlsr\ value that should not be neglected. Therefore
the final $R^2$ is defined as
\begin{equation}
   R^{2}=R^{2}_{\mathrm{spatial}}+R^{2}_{\mathrm{velocity}}~,
\label{eq:rsquare_sum}
\end{equation}
where $R^{2}_{\mathrm{spatial}}$ measures the difference of the spatial  
component between the observation and the model; $R^{2}_{\mathrm{velocity}}$
measures the difference of the observed \vlsr\ and that predicted in
eq.~(\ref{eq:vlsr}). The weightings for the two $R^{2}$ components are set 
to~1, as we suggest that both spatial and \vlsr\ information are equally 
important.

Figure~\ref{fig:4models} shows the schematic view of the four different 
possible scenarios for the model.
Since we assume two jet patterns, we have to objectively consider all the 
cases in which they have different/similar apparent origins, and whether they 
have different/similar precessing directions.
The best fit parameters are obtained by maximizing $R^{2}$. 
The fitting results of the four scenarios are shown in Table~\ref{tab:model}, 
and the graphical illustrations for different variations of the model are 
shown in Figure~\ref{fig:modelC} to \ref{fig:modelC_opp}.
Note that the small fitting errors listed in Table~\ref{tab:model} only 
include the statistical variance from the fitting routine, and they give a 
68\% confidence level (1 sigma).
Jet~1 is easily identified because many maser features can be
found along nearly the whole jet, from the driving source up to the northern 
and southern tips, 
as shown in the figures. 
Statistically the fitting result for jet~1 is the most reliable
among those presented in this paper, with $R^{2}=0.86$. 
The systemic velocity of the driving source is $\sim$47\kms, and the 
precession period is $\sim$56~years. Jet~1 has a counter-clockwise
precessing direction (note that in this paper, ``clockwise'' or 
``counter-clockwise'' refers to the precessing direction observed from the 
northern side of the jet axis).
On the contrary, in jet~2, 
few maser features appear in the region near to its driving source as 
most of them are located at the tips. In this case, it is numerically 
possible to fit several variations of the
model to jet~2. 

From the model, jet~1 can reproduce the gradual change of the
line-of-sight velocity of the maser features distributed along the helix 
curve, which is a typical pattern due to precession. For jet~2, there
is not enough data, so even though we
obtain reasonable values of $R^{2}$, the numerical results obtained by 
fitting might not 
be very accurate. Nonetheless, the existence of jet~2 should not be rejected 
solely because of the relatively small number of maser features. In fact, 
the inclusion of jet~2 in our model is necessary to explain a clear 
characteristic of the observations:
in region I and II of Figure~\ref{fig:modelC}, the mixing of maser features 
with different line-of-sight velocities is well explained by the (apparently) 
overlapping effect of the two jets. The same thing happens for the two other
scenarios as illustrated in Figure~\ref{fig:modelC_one} and \ref{fig:modelC_opp}. 
In addition, the 
deviation of vector 
direction, as mentioned previously, can be interpreted by having two jets 
with different driving sources spatially close to each other as demonstrated 
by our model (be it resolvable or not). 
More support for our hypothesis is given in the three position-velocity 
diagrams (figure~\ref{fig:dec_vlsr} to \ref{fig:dec_vdec}, jet~$2a$ is 
chosen as an example as it gives the best $R^{2}$). It is observed that
a single jet is not enough to describe the wide velocity 
distribution of the \water\ maser features at different positions. 
Figure~\ref{fig:modprop} shows another qualitative comparison 
between the model predicted proper motions and the observed proper 
motions of the \water\ maser features. The model can correctly describe not 
only the spatial and $V_{\mathrm{LSR}}$ distribution (indicated by the 
colors), but the proper motions of the features as well (indicated by the 
arrows). 

The best fit result for jet~2 is obtained when it is assumed that both
jets are precessing in a counter-clockwise direction but originate from 
two resolvable driving sources (jet~$2a$ in Table~\ref{tab:model});
in this case $R^{2}=0.73$. The systemic velocity and precession period for 
jet~$2a$ are
$\sim$61\kms\ and $\sim$73~years respectively, which are clearly different 
from those of jet~1. The second plausible case is that jet~2 precesses in
the counter-clockwise direction and originates from the same 
point as jet~1  (jet~$2b$ in Table~\ref{tab:model}). It does not necessarily
mean that 
the two jets have physically the same driving source, but it represents 
the scenario in which the two sources are spatially unresolvable. Under this 
setting $R^{2}=0.54$, this is the lowest among all three trials. 
The systemic velocity is $\sim$54\kms\ 
and the precession period takes the value of $\sim$53~years, which is very 
close to that of jet~1. The third possible case is significantly different from
the above two, jet~2 is now assumed to be precessing in an opposite
direction to jet~1, i.e. in the clockwise direction, and they have two 
resolvable driving sources (jet~$2c$ in Table~\ref{tab:model}). Even though
the precessing direction is different, this model is able to produce a 2D
projection well fitted with the observation. Here $R^{2}=0.70$, which is
similar to the first case; the systemic velocity and precession period for 
jet~$2c$ are 
$\sim$56\kms\ and $\sim$95~years respectively. This case has the longest
precession period. Logically, there should be a fourth case, which corresponds
to scenario (d) in Figure~\ref{fig:4models}. In this case jet~2 
has opposite precessing direction to jet~1 but they share the same
origin point (i.e. unresolvable driving sources). However, it is found
that the fitting result is untenable and
so is not discussed further.

There are still some discrepancies between the model and the observations 
even if we fit two jets on the position-velocity diagrams,
and the model cannot handle the few maser features
that are scattered in the area near to the driving sources.
They could hint at an as yet undiscovered equatorial outflow. 
Such an outflow associated with water fountains has been predicted by 
\citet{imai07iaup} to explain the existence of low velocity maser 
components in W\,43A and IRAS\,18460$-$0151, together with the 
narrow profile of the CO~$J=3-2$ emission line in \target\
\citep{imai09pasj}.  
Unfortunately, with only a few features identified in this area, quantitative 
analysis for this possible new component is out of the question for now, but 
at least we are confident that the masers are not lying on the proposed jets.  
If there are other kinematical components in addition to the jets, then 
the current model is not able to give very accurate answers. 
Nonetheless, the double-jet model offers one possible explanation
for the observational results with a statistical basis.  
A discussion about the three different choices for jet~2 is given
in Section~\ref{ssec:twojets}.

\subsection{Other Possibilities}
\label{ssec:bi}

A bi-cone model, consisting of two jet cones connected by the tips, 
can also be tested.
The bi-cone shape is typical in planetary nebula \citep{kwok08iaup}. 
If we assume water fountains play an important role in the shaping process
as described in the introduction, the
jet might somehow have a morphology similar to that of a planetary nebula.
This model is simple in construction, and no unusual assumption is required. 
A schematic view of this 
model is given in Figure~\ref{fig:bicone}. If the jet axis is tilted as shown,
then in the northern cone, the side
closer to the observer has a larger red-shift than the opposite side. A 
similar argument can be applied to the blue-shifted southern cone. Therefore,
the wide-angled jet cones are able to explain why there are maser features 
with significantly different \vlsr\ appearing in the same region (e.g. region~I
and II) along the line-of-sight direction (i.e. the second characteristic of 
the outflow from \target\ listed in Section~\ref{ssec:spatial}). However, 
this model cannot explain the continuously 
changing \vlsr\ along the jet, the collimated arc-shaped distribution of the 
maser features, and the formation of the void regions (i.e. the first and 
third characteristics) and is therefore not considered further. 

An outflow consisting of multiple outbursts could be another possible 
explanation to the maser distribution. However, to produce such large 
variations in the \vlsr\ as shown in Figure~\ref{fig:maserC}, either the 
velocity of an individual outburst is very different from one to the other, or
the jet axis is continually changing. It is difficult to explain such a physical
condition, and we therefore consider such a model unfeasible.

\section{Discussion}
\label{sec:dis}

We have shown in the model that the maser distribution can be explained by
two precessing jet patterns, but several questions remain unresolved. 
What is the origin of such jet patterns? What causes the spiky spectral 
profile of \target? Such profiles are not commonly observed in water fountain sources but
are often found in YSOs. How can we deduce the evolutionary status of this
object? We have mentioned another water fountain with a precessing jet, W\,43A, 
what are the similarities and differences between \target\ and 
W\,43A? These questions are discussed in this section.

\subsection{How to Produce Two Jet Patterns?} 
\label{ssec:twojets}

In Section~\ref{ssec:pre}, we demonstrated that the distribution and kinematic
properties of the maser features can be explained by two precessing jet 
patterns. The scenario for jet~1 is quite clear, and for jet~2 three
possible cases are investigated. Each of them gives a different set of best 
fit parameters (see Table~\ref{tab:model})
and naturally lead to different physical interpretations. 
The pros and cons of each case are discussed before we come to a conclusion.  

Jet~$2a$ in Table~\ref{tab:model}, for which $R^{2}=0.73$, gives the best fit 
model for jet~2. It is assumed in this model that jet~2 precesses in a 
counter-clockwise direction, similar to
jet~1, but they have two resolvable driving sources separated by $\sim$16~mas
(equivalent to 64\,AU at 4\,kpc). A possible scenario is that the ``two''
jets are actually formed by one single source but at different instants, and
the source itself has a secular motion across the sky, moving from the 
position of jet~$2a$'s driving source to that of jet~1 .
This argument can explain why jet~2 has few maser features along the jet 
path other than at its tips. If the outflow is episodic, such that it will stop 
and resume after some time, the maser features at the tips of jet~2 mark the 
end of the previous ejection. The estimated ``age'' of this jet 
($\sim$30~years) then reveals the time passed since this ejection. Jet~1 has
a dynamical age of $\sim$19~years, which means at least 19 years ago the 
driving source was located at (or almost at) the current position, which 
is the fitted center of jet~1. In that case the driving source has taken 
less than 11 years to travel 64\,AU (if only the projected 2D motion is considered), 
and its secular motion velocity is estimated to be $\sim$27\kms, if
constant. The deviation in the systemic velocity might 
favor the fact that the secular velocity is in fact changing. 
In that case the driving source is probably accelerated or has an orbital
motion. Nonetheless,
the idea of a positional shift for the driving source does not contradict
previous observations of \target. 

The 3D velocity difference between jet~1 and jet~2 might also suggest
that they represent two separate expulsion events.
When the driving source is moving, the jet orientation with
respect to Earth can vary. Therefore, the change in the projected jet axis
(shown in Figure~\ref{fig:modelC}) between the events is not surprising.
Figure~\ref{fig:moving} shows a schematic diagram of this moving-source model.
The maser features that produce jet~2 are excited by the previous ejection, 
and as time passes they move to the tip of the jet leaving almost no traces
along the path (the ``void'' regions of Figure~\ref{fig:maserC}). 
The OH~maser,
whose position is shown in Figure~\ref{fig:maserC},
is excited in the circumstellar envelope that moves along with 
the central star, and it provides an approximation of 
the latest position of the central star, which is consistent with our model.
Jet~1 is ``newly formed'' and
the ejection is probably still on going, therefore we can see 
the water maser features along the jet path, with the origin of jet~1 
being the current position of the driving source.
However, since the number of maser features associated with
jet~2 is relatively small and
they are concentrated at the tips, the fitting results have a larger
uncertainty comparing to that of jet~1. 
This could be one of the reasons why jet~1 and jet~$2a$ have different 
apparent precession angles and periods, while it is assumed that they are from the 
same source.

Jet~$2b$ in Table~\ref{tab:model} is the second possible case for modelling 
jet~2. It has almost the same construction as jet~$2a$, but the number of
apparent driving sources is different. In this scenario, jet~2 is also assumed 
to have the same precessing direction as jet~1 but this time both jets appear 
to originate from the same point, i.e. their driving sources are spatially 
unresolvable. Though the construction is similar, it is noted that the goodness
of fit of jet~$2b$ ($R^{2}=0.54$) is much worse than that of jet~$2a$. It seems
that the choice of having two resolvable/unresolvable driving sources is a 
crucial issue. Nonetheless, the same hypothesis 
as proposed for jet~$2a$ can be applied here, and this time the problem
of the discrepancy in precession period and precession angle does not 
occur as $T\sim56$~years, $\alpha\sim28$~degrees for jet~1, and 
$T\sim53$~years, $\alpha\sim22$~degrees for jet~$2b$. The only 
difference between this scenario and that discussed above is the assumption that 
the proper motion of the driving source is too slow to be observed in this 
case, and thus it appears to be stationary.

Jet~$2c$ in Table~\ref{tab:model} is the third possible case, with 
$R^{2}=0.70$. The appearance of the fitted 2D curve looks similar to 
jet~$2a$, but the two cases (i.e. jet~$2a$ and $2c$) are actually quite 
different, as seen from the orientation of the jet axes 
(Figure~\ref{fig:modelC} and \ref{fig:modelC_opp}). Here, jet~2 is
assumed to be precessing in the opposite direction to jet~1, and the driving
sources are spatially resolvable. 
In the previous model cases of jet~$2a$ and $2b$, there is in fact only one 
jet with one driving source but it produces the pattern of two jets.
This idea does not apply here 
because of the opposite precessing direction. Under this assumption we 
require two independent jets. This is kinematically possible and this 
time the missing maser features of jet~2 might be explained by the uneven
distribution of ambient matter. The main problem is whether it
is really physically possible to have two systems, of almost identical 
evolutionary status, launching jets in such 
close proximity. The two driving sources are separated by $\sim$17\,mas 
(about 68\,AU at 4\,kpc), the radial separation is unknown. 

The combination of jet~1 and jet~$2a$ is the most plausible case
from both a statistical and physical point of view. With the current data, we 
cannot expect this to be a unique model, but we believe it can satisfactorily
explain the observational 
characteristics of \target. In Section~\ref{ssec:pre},
we mentioned that it is possible to explain all the jet 
characteristics
by using two independent driving sources. That would mean the two jet patterns 
are really formed by two different systems. It is clear that this construction
gives us a lot of freedom in our modelling and the kinematical difference 
between jet~1 and jet~2 would be naturally explained.
Nonetheless, this idea would imply that there are two ``water fountain''
objects at the same spot on the sky. We cannot totally reject this possibility,
but given the fact that there are only 13 water fountains so far discovered 
across the sky, the probability of having two such objects at one site is very 
small. Hence we suggest that the two jet patterns are more likely coming from 
a single but moving source. Note that even though jet~$2a$ and $2c$ have
very different configurations, they both have their best fitted centers at
similar locations. We believe this provides a constraint on the 
appropriate distance ($\sim$16\,mas) between the two apparent driving sources,
if they are resolvable. Finally, in our fitting, although the
driving source(s) are assumed to be stationary for simplicity, this assumption
does not contradict the above moving-source idea because the 
travelling speed of the driving source is slow compared with the jet velocity.
A detailed kinematical model, which includes a travelling source,
will be part of our future work.

\subsection{The Feasibility of Having Two Resolvable 
            Apparent Sources} 
\label{ssec:kstest}

From the above discussion, the models for jet~$2a$ and $2b$ 
(i.e. one single jet forming the pattern of two jets)
are seen to be most plausible. In the former case,
the two jets are apparently driven from two resolvable sources in order to
obtain reasonable $R^{2}$ values in the fitting.
The remaining issue is whether this prediction of having two 
visually resolvable radiant points is feasible or not. 
A quantitative analysis is helpful for this matter.

The straight lines in Figure~\ref{fig:lines} represent the extension of each 
proper motion vector as shown in Figure~\ref{fig:proper}. For ballistic motion,
if all the vectors originate from a single point, the straight 
lines should converge to one point. This happens when a single source model is
being considered, or a model with two unresolvable sources (jet~$2b$). 
The other case is of course, 
a model with two resolvable sources in which the straight lines should 
intercept at two distinct points. In reality, owing to the uncertainty of the 
maser positions, the observed proper motion for each maser feature has a
measurement error and the extensions do not 
converge 
to one or two points only, instead the interception points for every pair of 
lines are distributed in an extended region around the real radiant point(s). 
The pattern in Figure~\ref{fig:lines}(a) is therefore expected. However, this 
figure alone is inconclusive as the pattern can be generated by a 
single radiant point or two radiant points 
that are visually close to each other (e.g. jet~$2a$). 

We will build two simulation models, one with a pair of unresolvable 
driving sources (single apparent radiant point) and another one with a pair of 
resolvable driving sources, to compare them with the observations. 
Figure~\ref{fig:lines}(b) shows the interception points for every two lines
in Figure~\ref{fig:lines}(a). The distribution of the points hints at the 
possible locations in which the dynamical centers could be found
(i.e. the dense region at the center). The elongated
pattern for the dense region is due to the bipolarity of the 
system (most of the maser features are concentrated along the north-south 
direction). As mentioned in the previous paragraph, one of the main 
reasons for the dispersion of the points is the error in the measured proper
motion, this has to be taken into account for the models. The driving 
source for jet~1 is chosen as the single-radiant-point model due to its high 
degree of reliability as mentioned in the previous discussion. For the other
model with two resolvable origins, we use the driving sources of jet~1 and
jet~$2a$ as the radiant points.
In order to do a fair comparison,
the same 54 maser features as listed in Table~\ref{tab:proper} are used for 
generating the ``artificial'' proper motions. The steps for one simulation of
a model are as follows:
\begin{enumerate}
\item The maser features with index~1 in Table~\ref{tab:proper} are used,
      for simplicity, we group these coordinates as set~A. 
\item Assuming the features move along straight lines from the chosen 
      radiant point(s), we can calculate the new positions of the features half
      a year later. This new set of coordinates are referred to as 
      set~B. Therefore, each member in set~B represents the predicted location
      of each corresponding member in set~A half year later, under ideal
      conditions.
\item Table~\ref{tab:spots} shows that the positional uncertainty of the 
      maser features is $\sim$0.1\,mas in all the epochs.  Therefore, in order 
      to reproduce the positional uncertainties in the real 
      observables, random ``errors'' following a 2D Gaussian distribution 
      with $\sigma=0.1$\,mas are added to all the coordinates in set~B. 
\item Now we have two epochs of data, set~A and B. We can calculate the proper
      motions and make a map similar to Figure~\ref{fig:lines}(b). The data
      points here are grouped in set~C.
\end{enumerate}
The above steps are repeated 1000 times for both models.

Figure~\ref{fig:points}(a) shows one of the simulation results (set~C) for the 
model with unresolvable driving sources, and Figure~\ref{fig:points}(b) shows 
the result for the model with resolvable driving sources. 
The model for the resolvable case is able to
reproduce the ``inverted triangle'' feature found in the observation data 
(indicated by the dotted black boxes in Figure~\ref{fig:lines}(b) and 
\ref{fig:points}(b)), while the other fails to
do so. To produce this feature, two different trends of proper motions are 
needed and therefore only the resolvable case could fulfil this criteria.

For further analysis, each time set~C is
obtained, it is compared with the observed data (Figure~\ref{fig:lines}(b)) by 
the Kolmogorov-Smirnov test \citep[or K-S~test, see for example,][]{lupton93}. 
The ``bins'' of the K-S~test are assigned in such a way that each bin 
represents different distances between a data point to the radiant point(s),
with a resolution of 1\,mas~bin$^{-1}$. 
The averaged p-value, which represents the 
null hypothesis of the two sets of data being ``the same'', has a 
value of 5\% for the unresolvable-driving-sources model, while for the
resolvable case, the value is 51\%. We believe that the unresolvable case 
can be rejected with a 95\% confidence level.

The above results are therefore consistent with 
the idea of having two resolvable radiant points, and the jet~1 and jet~$2a$ 
combination remains as a plausible explanation to the observed maser distribution,
even though some of the parameters require confirmation.
Finally, it should be noted that the crossing of the two jets in region I and 
II of Figure~\ref{fig:modelC} is just a visual effect, the jets actually 
do not meet as seen from our 3D version of the model. In fact, if the jets 
really collide with each other, the maser kinematics in the colliding regions 
would be more complicated than that observed.

\subsection{Evolutionary Status and a Comparison with W\,43A}
\label{ssec:evo}

\begin{figure}[ht]
   \centering
   \includegraphics[scale=1]{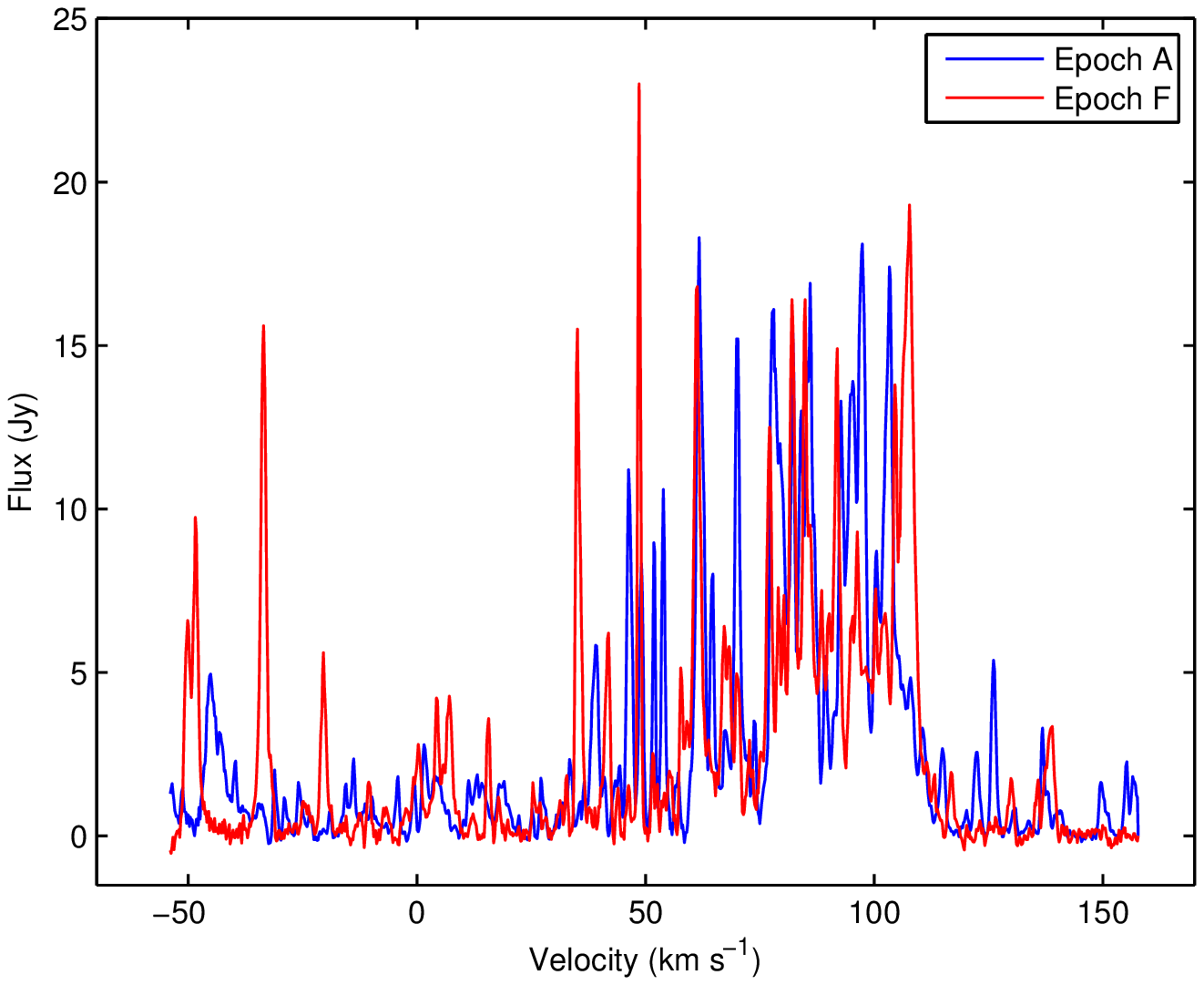}
   \caption{\water\ maser spectra of \target\ in epoch~A (2008~April~21) and 
            F (2009~May~19) of the VLBA observations. 
            The spectra are created by the MIRIAD task IMSPEC, and the 
            integration regions are selected in a way such that all the maser 
            features are included. The spectrum for epoch~A is displayed in 
            blue while that for epoch~F is in red. The spiky profile is 
            one of the characteristics of \target. Emission peaks are found
            in the range of $-50$\kms\ to $150$\kms\ for all the epochs, but
            a significant change in the profile is noticed over the
            observation period.}
   \label{fig:afspectra}
\end{figure}

\begin{figure}[ht]
   \centering
   \includegraphics[scale=1]{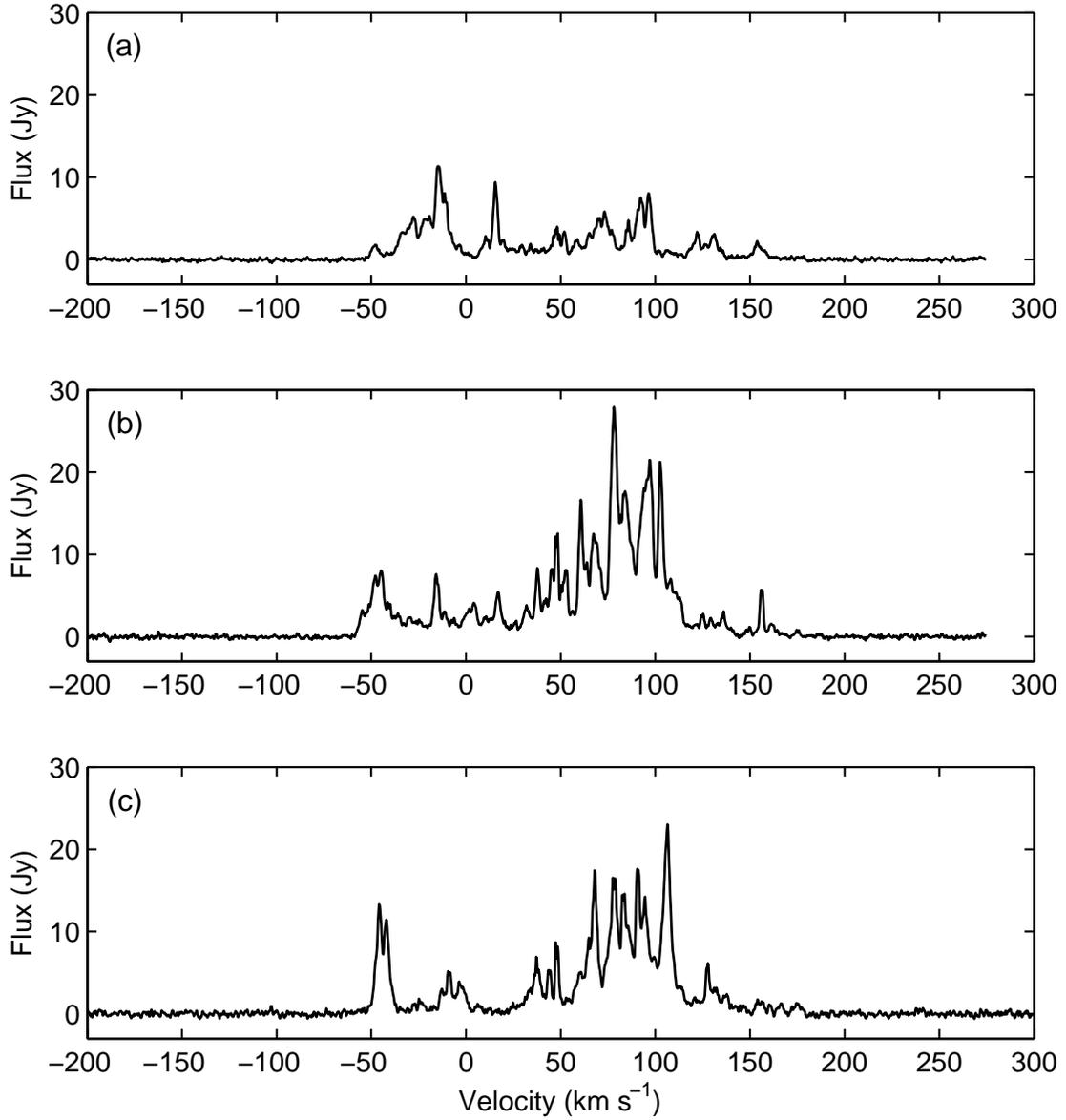}
   \caption{\textsl{(a)}: \water\ maser spectrum of \target\ taken by the 
            NRO~45\,m telescope on 2006~April~20, the observation details 
            can be found in \citet{deguchi07apj}.
            \textsl{(b)} and \textsl{(c)}: Same as (a), but taken on 
            2008~April~30 and 2010~April~2, respectively. 
            }
   \label{fig:nrospectra}
\end{figure}

\begin{figure}[ht]
   \centering
   \includegraphics[scale=1]{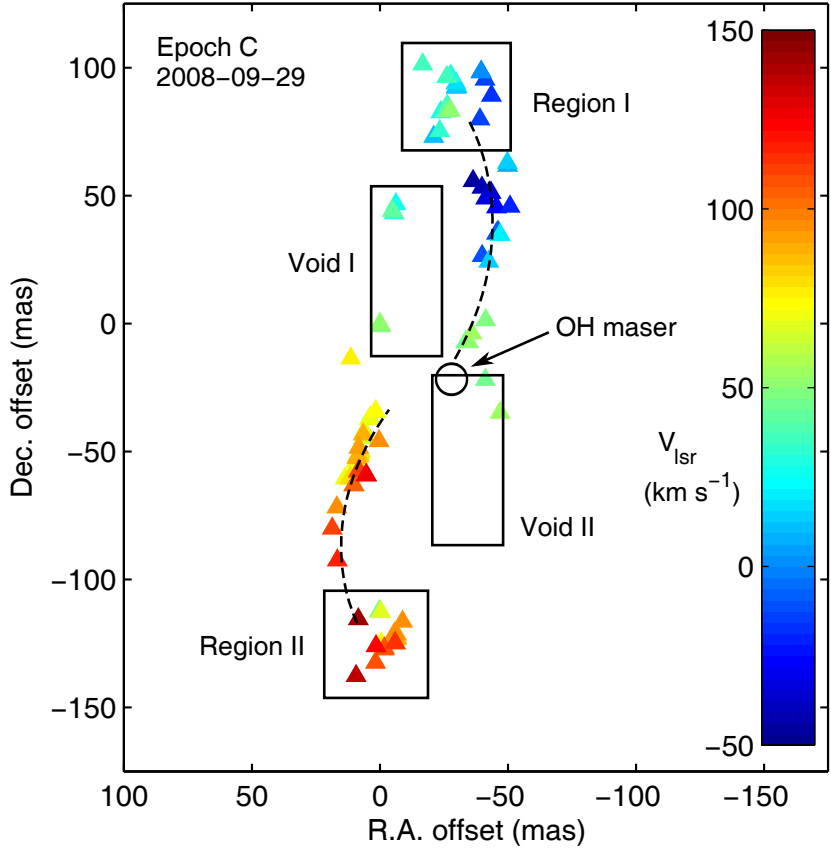}
   \caption{Spatial distribution of \target\ \water\ maser features in 
            Epoch~C.
            Since the appearance looks very similar for all epochs, 
            only this one is shown here. In Epoch~C, 
            most of the features with measured proper motions were detected.
            (shown in Figure~\ref{fig:proper}). 
            Each maser feature is represented by
            a filled-triangle and the color denotes its line-of-sight velocity
            according to the scale of the color bar. 
            Most of the maser features are lying on the two dotted arcs, but
            in \textsl{Void I} and \textsl{Void II} only a few of 
            them are found.
            \textsl{Region I} and \textsl{Region II} show clusters of maser 
            features with wide ranges of line-of-sight velocities 
            ($-40$\kms$\,<\,$\vlsr$\,<40$\kms\ and 
            $70$\kms$\,<\,$\vlsr$\,<150$\kms,
            at the northern and southern end of the 
            structure, respectively). 
            The map origin is set at the chosen reference feature.
            The black circle indicates the position of the OH maser emission
            \citep{imai08evn}.   
            }
   \label{fig:maserC}
\end{figure}

\begin{figure}[ht]
   \centering
   \includegraphics[scale=1]{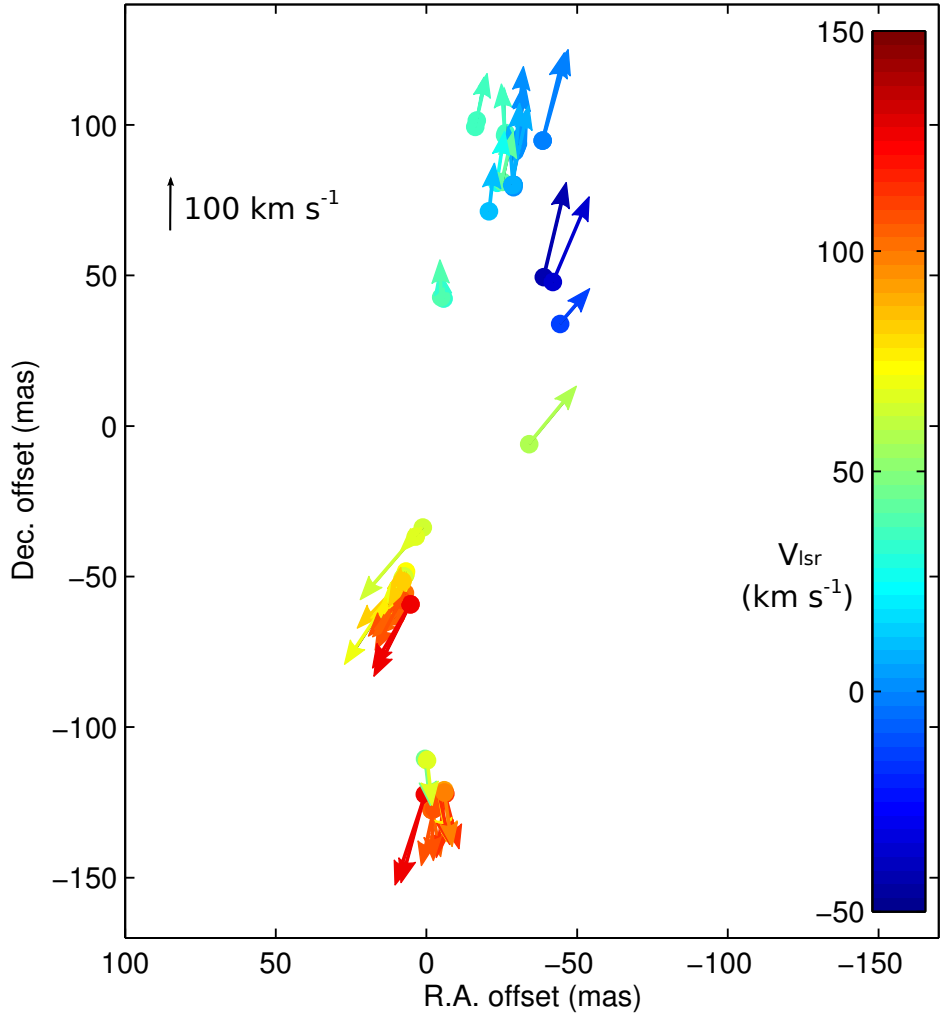}
   \caption{Proper motions of 54 \water\ maser features in \target\ which 
            can be traced in any 3 consecutive epochs. The filled-circles
            represents the data in Table~\ref{tab:proper} with index~1.
            }
   \label{fig:proper}
\end{figure}

\begin{figure}[ht]
   \centering
   \includegraphics[scale=1]{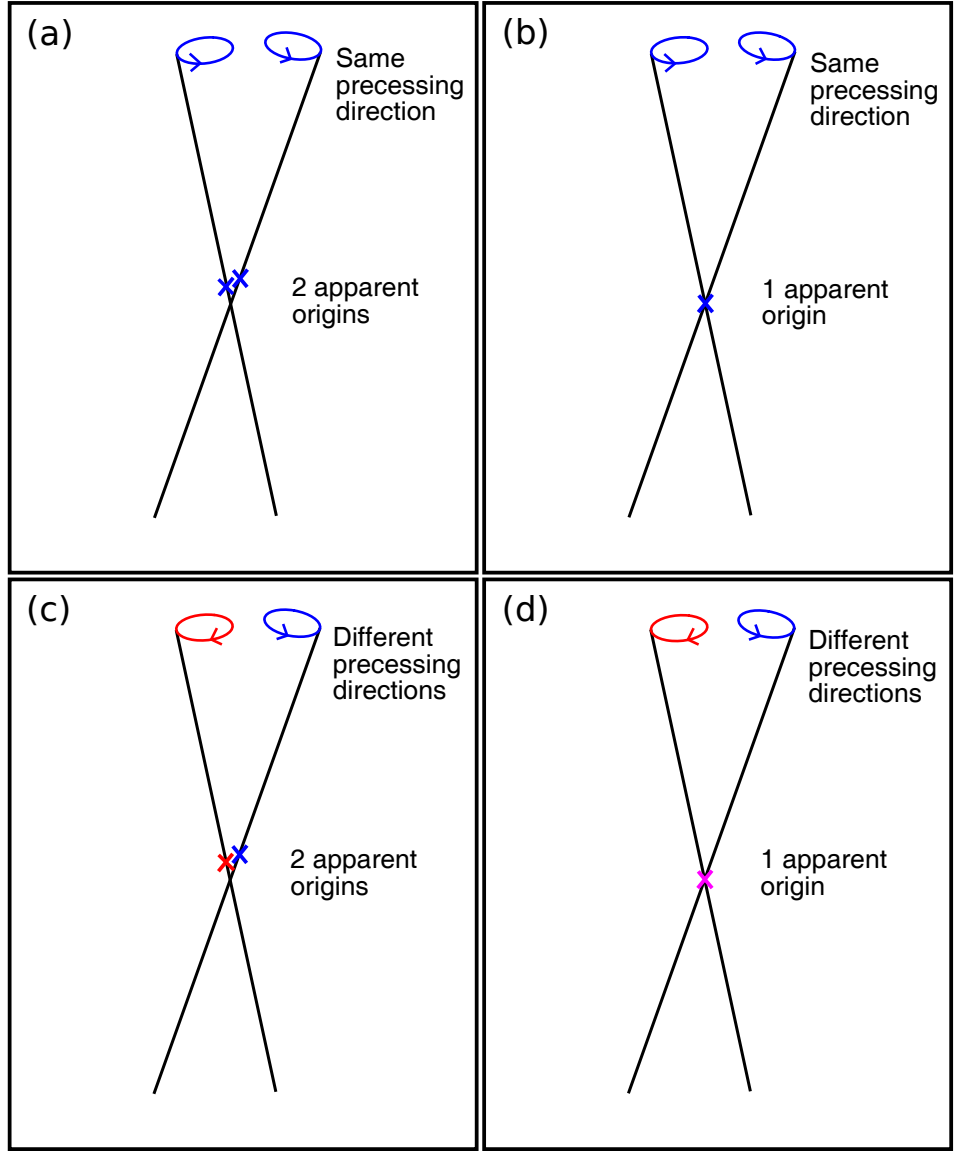}
   \caption{Four different possible scenarios for the precessing jet model. The
            main differences between each of them are the choice of precessing
            directions and the number of apparent jet origins.} 
   \label{fig:4models}
\end{figure}

\begin{figure}[ht]
   \centering
   \includegraphics[scale=1]{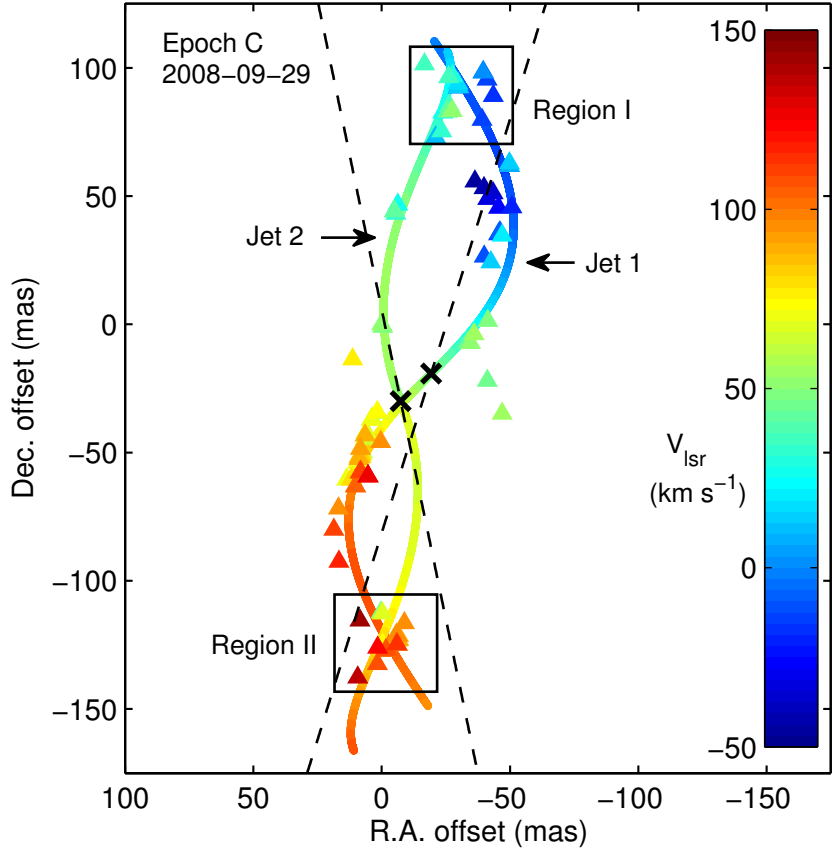}
   \caption{Illustration of the double precessing jet model with two
            resolvable driving sources, superposed
            on Figure~\ref{fig:maserC}. Since for 
            every epoch the fitting procedure is the same, here only the
            fitted curve and maser data of epoch~C are shown as an example. 
            The two jets have the same anti-clockwise precessing 
            direction. The black dotted
            lines show the projection of the jet axes on the sky plane.
            Jet~1 is the dominant jet with more \water\ maser features lying 
            on it, while jet~2 is more subtle.
            The two driving sources (represented by
            two black crosses) are separated by
            $\sim$16\,mas.} 
   \label{fig:modelC}
\end{figure}

\begin{figure}[ht]
   \centering
   \includegraphics[scale=1]{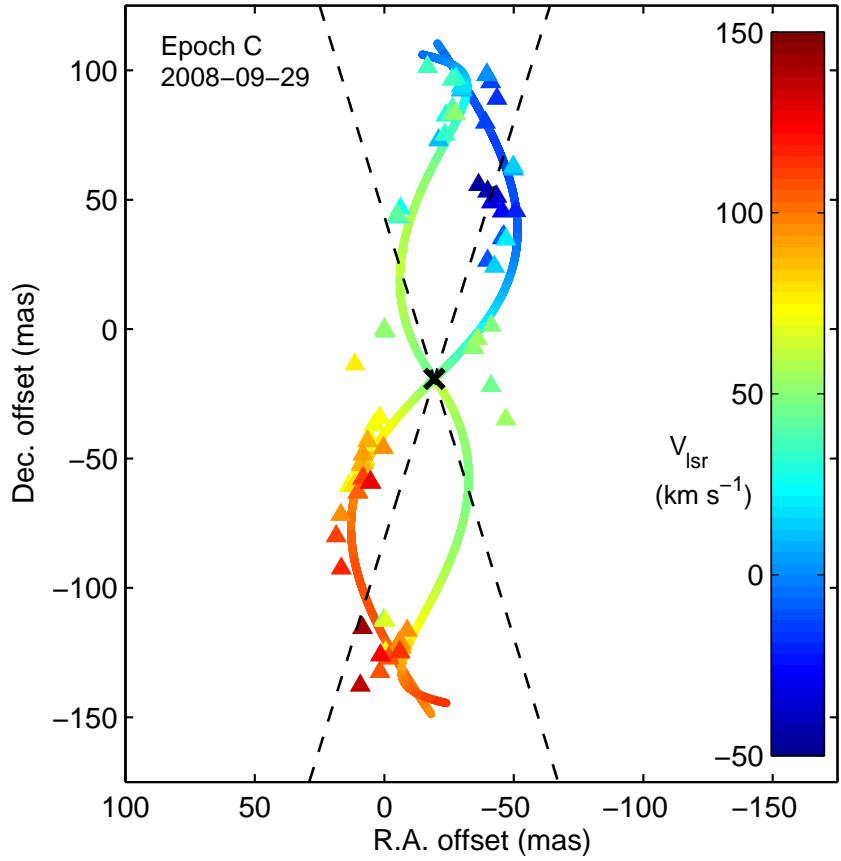}
   \caption{Same as Figure~\ref{fig:modelC} but for the double precessing 
            jet model with one
            driving source (i.e. the driving sources, even if there are more 
            than one, are assumed to be unresolvable). 
            The two jets have the same anti-clockwise precessing 
            direction.
            } 
   \label{fig:modelC_one}
\end{figure}

\begin{figure}[ht]
   \centering
   \includegraphics[scale=1]{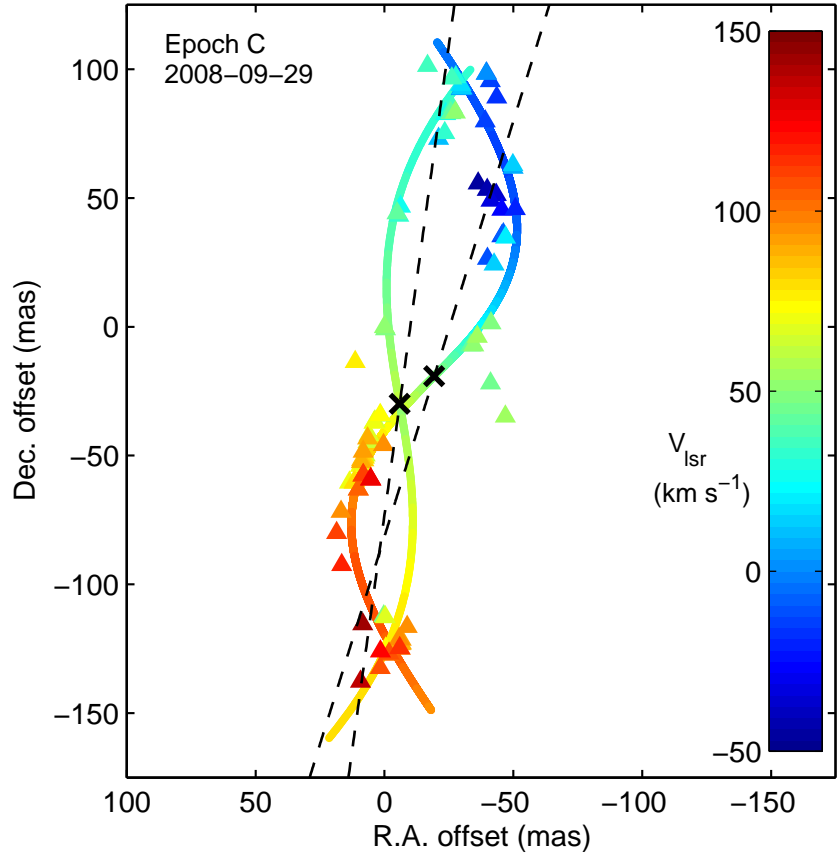}
   \caption{Same as Figure~\ref{fig:modelC} but the 
            two jets have opposite precessing directions.
            The two driving sources (represented by
            two black crosses) are separated by
            $\sim$17\,mas.
            } 
   \label{fig:modelC_opp}
\end{figure}

\begin{figure}[ht]
   \centering
   \includegraphics[scale=1]{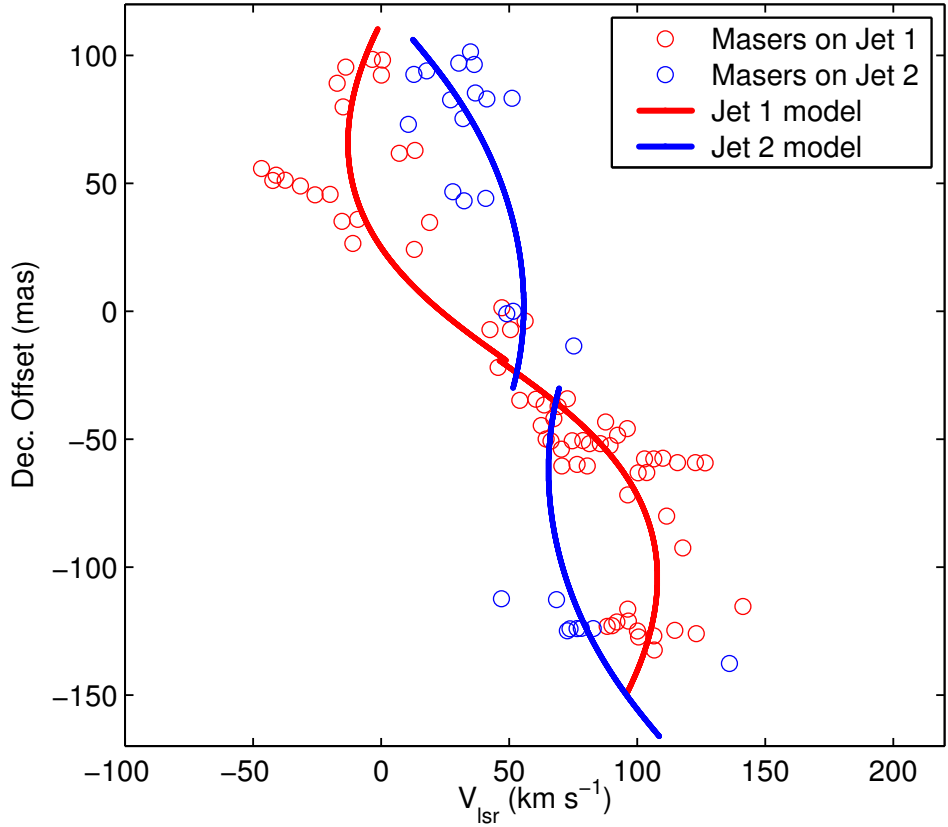}
   \caption{Declination offset of the maser features in epoch~C against the
            corresponding line-of-sight velocities, together with the 
            predicted values from the model fitting as shown in 
            Figure~\ref{fig:modelC}. The data points and the 
            model curve for jet~1 are represented by open-circles and line
            in red, while those for jet~2 are in
            blue. The declination of the reference maser feature is zero.}
   \label{fig:dec_vlsr}
\end{figure}

\begin{figure}[ht]
   \centering
   \includegraphics[scale=1]{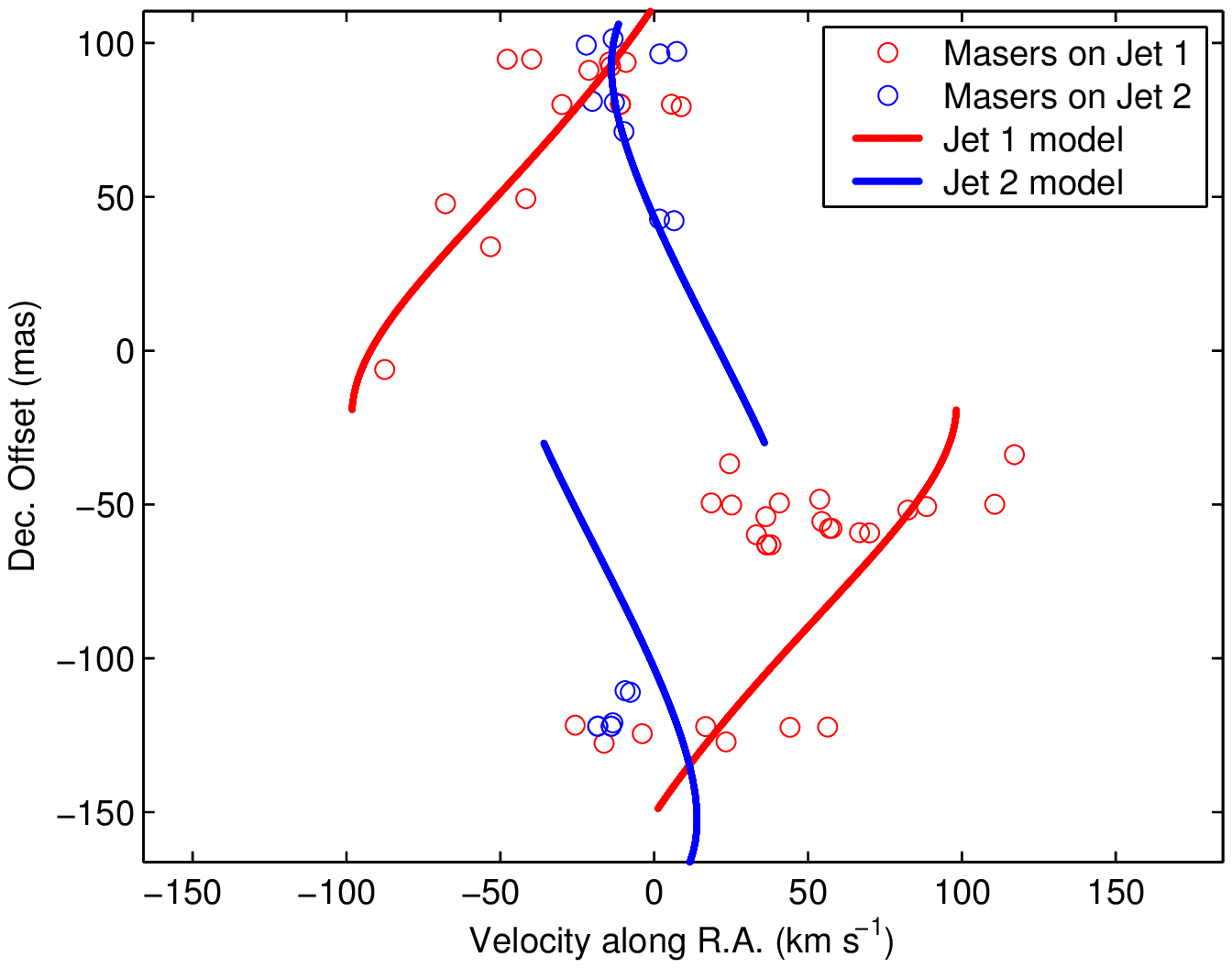}
   \caption{Same as Figure~\ref{fig:dec_vra} but for the declination of the 
            maser features in Figure~\ref{fig:proper} against 
            the corresponding proper motion velocities along the R.A. 
            direction, together with the 
            predicted values from the model fitting as shown in 
            Figure~\ref{fig:modelC}.}
   \label{fig:dec_vra}
\end{figure}

\begin{figure}[ht]
   \centering
   \includegraphics[scale=1]{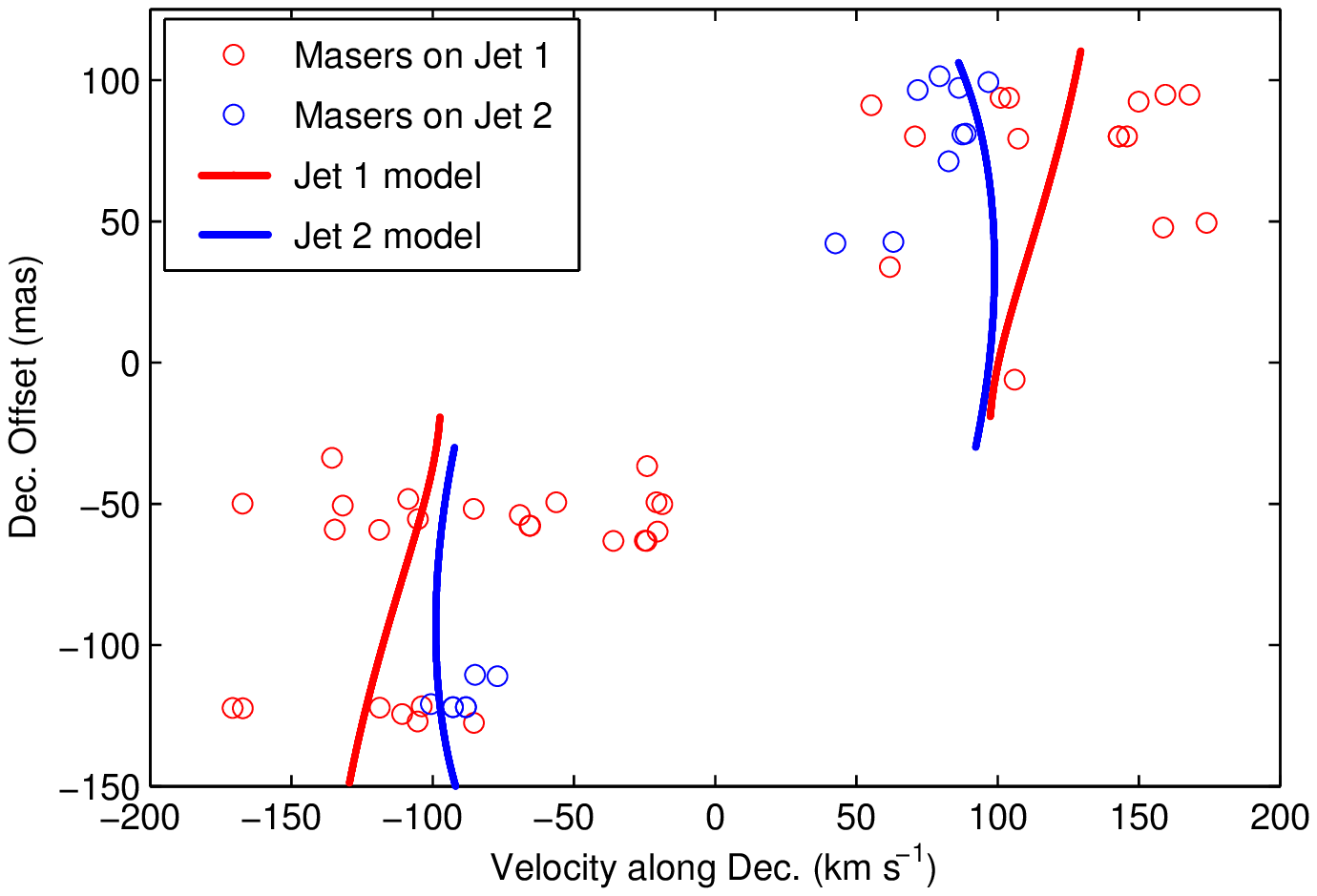}
   \caption{Same as Figure~\ref{fig:dec_vra} but for the declination of the 
            maser features in Figure~\ref{fig:proper} against 
            the corresponding proper motion velocities along the Dec. 
            direction, together with the 
            predicted values from the model fitting as shown in 
            Figure~\ref{fig:modelC}.}
   \label{fig:dec_vdec}
\end{figure}

\begin{figure}[ht]
   \centering
   \includegraphics[scale=1]{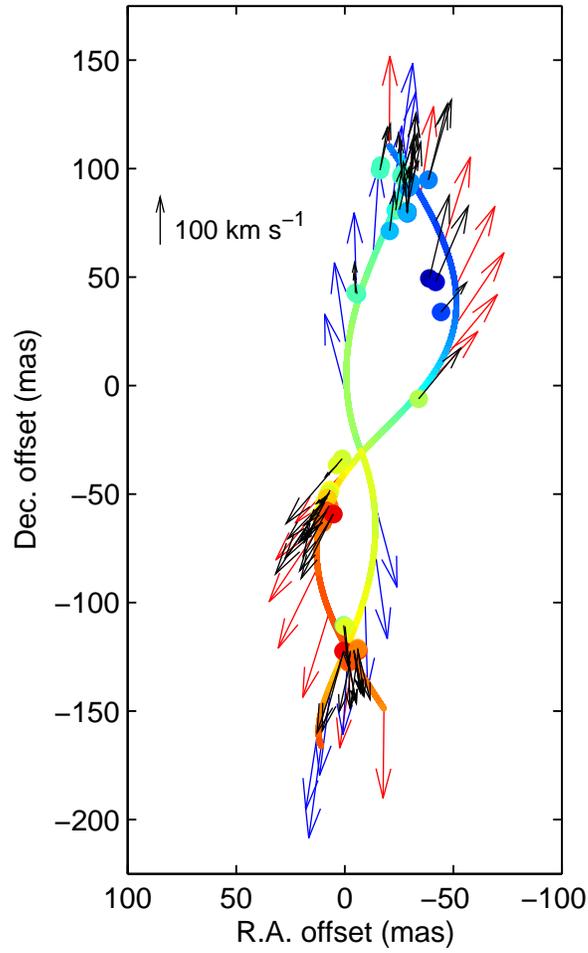}
   \caption{A comparison between the double-jet model in 
            Figure~\ref{fig:modelC} 
            and the observed proper motions of \water\ maser features. 
            \textsl{Black arrows}: Observed proper motions of \target, 
            which are the same as those illustrated in 
            Figure~\ref{fig:proper}.
            \textsl{Red and blue arrows}: Model predicted proper motions for
            maser features lying on jet~1 and jet~2.}
   \label{fig:modprop}
\end{figure}

\begin{figure}[ht]
   \centering
   \includegraphics[scale=1]{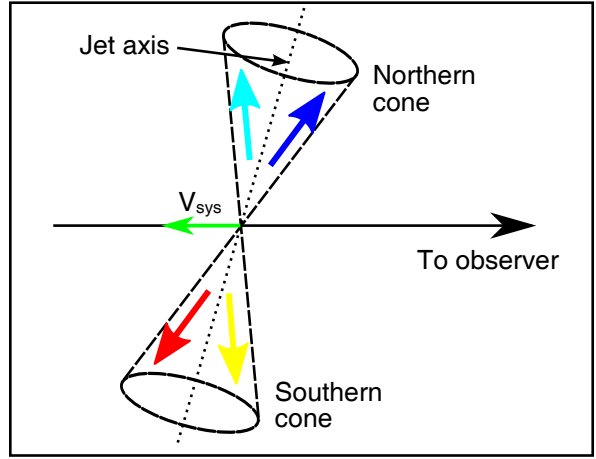}
   \caption{Illustration of the bi-cone model. The arrows show the outflow
            directions, and the colors indicate different line-of-sight 
            velocities of the corresponding parts of the cone, with reference 
            to the colorbar in Figure~\ref{fig:maserC}.
            }
   \label{fig:bicone}
\end{figure}

\begin{figure}[ht]
   \centering
   \includegraphics[scale=1]{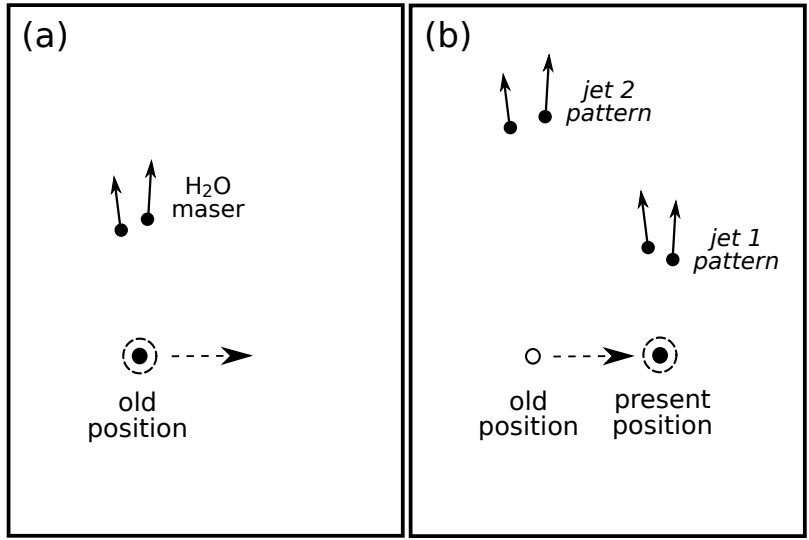}
   \caption{Illustration of the moving-source model which produces two jet
            patterns from one driving source only. 
            }
   \label{fig:moving}
\end{figure}

\begin{figure}[ht]
   \centering
   \includegraphics[scale=0.9]{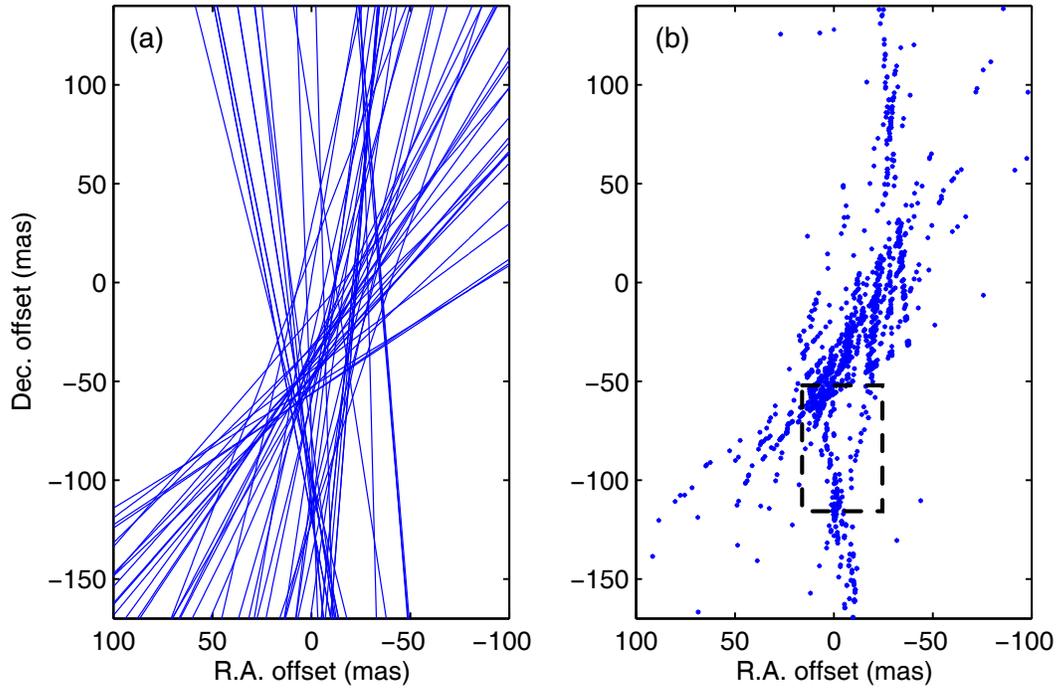}
   \caption{\textsl{(a)}: The extensions of the proper motion vectors. Each 
            straight line goes along the direction of a vector showing in 
            Figure~\ref{fig:proper}.
            \textsl{(b)}: The interception points between any pair of straight 
            lines as illustrated in (a). The driving 
            sources are most likely lying within the dense region.
            The dotted black box
            denotes the ``inverted triangle'' feature, which is also found in
            Figure~\ref{fig:points}(b).
            }
   \label{fig:lines}
\end{figure}

\begin{figure}[ht]
   \centering
   \includegraphics[scale=0.9]{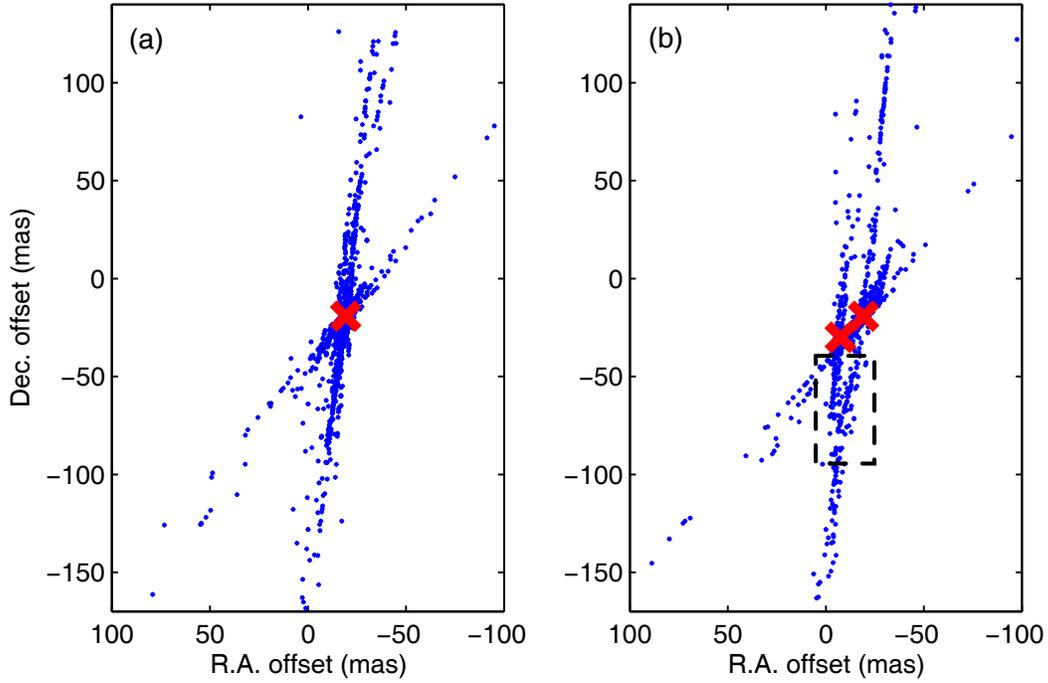}
   \caption{\textsl{(a)}: Similar to Figure~\ref{fig:lines}(b), but the data 
            are generated by a model with a single radiant point (to 
            represent the case where the two driving sources are unresolvable)
            The cross indicates the location
            of the assumed single radiant point, which has the coordinates
            of the driving source for jet~1.
            \textsl{(b)}: Similar to (a), but the model has two
            resolvable radiant points. The red crosses indicate the positions 
            of the driving sources of jet~1 and jet~$2a$. The dotted black box
            denotes the ``inverted triangle'' feature, which is also found in
            Figure~\ref{fig:lines}(b).
            }
   \label{fig:points}
\end{figure}

\target\ appears as a very red point source in both MSX and GLIMPSE images 
\citep{deguchi07apj}. Its position on the MSX color-color diagram suggests 
that it is an evolved star \citep[see, e.g. Figure~3 in][]{day10apj}. 
However, \target\ exhibits a spiky spectral profile as presented in 
Figure~\ref{fig:afspectra}; such a line profile, showing a number of peaks, 
is not typical in AGB and post-AGB stars, but is often found in star forming 
regions. Therefore, it is necessary to consider some basic characteristics of 
\target\ to understand its evolved star status before undertaking any further 
interpretation with this assumption. 

Even though many YSOs also have a spiky spectral 
profile, their velocity range is quite different. Water fountains such as 
\target\ and other examples mentioned in this paper exhibit 
\water\ spectra with a large velocity range ($\geq$100\kms), while for 
many YSOs the range is usually less than 50\kms\ 
\citep[see, for example,][]{elling10mnras,urquhart10aa}. 
Nonetheless, there are cases such as W\,49 \citep{mcgrath04apjs} and 
W\,51 \citep{genzel79aa}
in which the spectra show emission peaks spread across
100\kms\ or more, so only considering the velocity range
is not sufficient to determine whether \target\ is an evolved star
or a YSO. However, for YSO spectra, the high velocity
\water\ maser features are usually significantly weaker in flux than 
those at low velocity. This is not the case for \target\, 
where its spectra show bright peaks in the high velocity range 
(Figure~\ref{fig:afspectra}).

In addition, 
the NRAO VLA Sky Survey at 1.4\,GHz ($\lambda=21$\,cm) shows no continuum 
detection in the direction of \target\ \citep{condon98aj}, which means it 
has no significant H$_{\mathrm{II}}$ source there and hence the object is 
unlikely to be lying in a high-mass star forming region. 
The report on the detection of 1612\,MHz OH maser by \citet{sevenster97aas} 
also suggests that \target\ is not a low-mass YSO, as there are 
currently no such OH masers found in this type of object \citep{sahai07aj}.
Therefore, \target\ is likely to be an evolved star.


\target\ is the second example of a water fountain other than W\,43A 
in which a highly collimated precessing outflow is observed. 
\citet{imai02nature} state that the molecular jet of W\,43A has a 
precession period
of 55 years and an outflow velocity of about 150\kms. These values are
of the same order of magnitude as the jet(s) in \target\ 
(see Table~\ref{tab:model}).
It is unclear whether the similar kinematic
parameters are just a coincidence or that there is a physical reason behind 
it. However, 
it is worth looking at the similarities and differences between 
the two objects. The properties of the \water\ and OH masers are 
discussed below.

The \water\ maser features
of W\,43A are found in two well separated clusters on spatial maps, and they 
show a double-group spectral profile, which is common in 
water fountains as mentioned previously. 
Generally, \water\ maser emission is thought to be found
in a shocked region which is formed by the collision between an outflowing 
jet and the ambient gas of the star \citep{elitzur92}. For water fountains
such as W\,43A, this happens at the tips of the bipolar jets with high 
velocity, hence the double-group profile is produced, and the maser features 
appear to be collected in two isolated clusters on spatial maps.
In the case of \target, however, other than the spiky spectral profile, 
the maser features seem to trace the jet starting from 
the region close to the central star all the way up to the blue-shifted and 
red-shifted tips, as indicated in the spatial distribution of maser emission
(Figure~\ref{fig:maserC}). 
If the inner-wall of the ambient gas envelope is relatively close to the 
central star, \water\ maser excitation would occur in the vicinity of the 
star and move along with the jet penetrating into the gas envelope. The whole 
jet can then be traced out by maser emission. Furthermore, if the maser 
features 
have different line-of-sight velocities in various positions on the jet (as 
in the case of a precessing jet), a spiky profile with emission peaks of 
different velocities will be observed (Figure~\ref{fig:afspectra}). 

\citet{imai08evn} reports on VLBI observations of 1612\,MHz OH maser for
both W\,43A and \target. The spatio-kinematical structure of the OH
maser emission associated with W\,43A can be described as a spherically
expanding shell with an expansion velocity of $\sim$9\kms\ and radius of
500\,AU, assuming a distance of 2.6\,kpc. For \target, only 
a single emission peak was detected with line-of-sight 
velocity equal to 39.5\kms. If we assume that the OH maser detection is also 
part of a spherical expanding shell, then the expansion velocity is 
estimated to be $\sim$7.5\kms\ if the line-of-sight velocity for the central 
star of \target\ is taken as $\sim$47\kms\ ($V_{\mathrm{sys}}$ of jet~1). 
The ambient shells of the two 
objects then have similar expansion velocities. 

\section{Conclusions}
\label{sec:con}

Our VLBA observations reveal a double-helix pattern traced by 22.2\,GHz 
\water\ maser emission. Modelling results show that the pattern can be
fit by two bipolar precessing ballistic jets. Jet~1 has a period of 
$\sim$56~years which is similar to that of W\,43A \citep{imai02nature}, 
while jet~2 has a longer period of $\sim$73~years (if the case of jet~$2a$ is
adopted). The two apparent driving sources are separated by $\sim$16\,mas on
the sky plane, which is about 64\,AU at 4\,kpc.
From the proper motions of the maser features, the 3D jet velocities 
are found to be
$\sim$138\kms\ and $\sim$99\kms\ for jet~1 and jet~2 respectively, if they 
are assumed to be constant throughout the observation period. 
It is suggested from the above analysis that there is in fact one driving
source only, but its secular motion and the episodic outflow possibly produce 
the two jet patterns. 
\target\ is likely to be an evolved star
because of the wide velocity span of the \water\ maser emission,
the detection of a 1612\,MHz OH maser and its infrared characteristics. 
Further investigation is needed to 
search for the existence of an equatorial outflow as hinted by the maser
detection near to the driving sources but not lying on the jets. 
It is also worth obtaining additional interferometry data covering the 
high velocity components that have not been included in the current VLBA 
results. An enhanced model, which includes the possible proper motion of the
driving source, could be used in our future analysis. The discovery of 
\target\ shows that a precessing collimated jet is not just confined to 
W\,43A. These two objects have  
kinematical similarities, and it may mean that there are specific 
conditions for precessing jets to occur, but they are yet to be discovered.



\acknowledgments

This research was supported by grants from the Research Grants Council
of the Hong Kong Special Administrative Region, China 
(Project No.~HKU\,703308P, HKU\,704209P and HKU\,704710P) and the Seed 
Funding Programme for 
Basic Research of the University of Hong Kong (Project No.~200802159006). 
B.Y. acknowledges the support by the HKU SPACE Research Fund. 
H.I. has been financially supported by Grant-in-Aid for Young Scientists from 
the Ministry 9 of Education, Culture, Sports, Science, and Technology 
(18740109) as well as by Grant-in-Aid for Scientific Research from Japan 
Society for Promotion Science (20540234).
The National Radio Astronomy Observatory is a facility of the National Science
Foundation operated under cooperative agreement by Associated
Universities, Inc.






\appendix


\bibliography{ms}

\clearpage

\clearpage

\begin{table}[ht]
\caption{Parameters of the VLBA observations and data reduction for each 
         individual epoch. \label{tab:status}
        }
\scriptsize
\begin{tabular}{l@{\ \ \ }c@{\ \ \ }ccl@{\ \ \ }r}
\hline \hline
Observation & Epoch & &
& \multicolumn{1}{c}{Beam\tablenotemark{c}}
& \\
code & (yyyy-mm-dd) & $V_{\mbox{\scriptsize ref}}$ \tablenotemark{a} 
& Noise\tablenotemark{b} 
&  \multicolumn{1}{c}{(mas)} & $N_{\mbox{\scriptsize f}}$\tablenotemark{d} 
\\ \hline 
BI37A \dotfill & 2008-04-21 & 59.8 & 1.1  
   & 1.81$\times$1.05, $-1\fdg$5 & 143 \\ 
BI37B \dotfill & 2008-05-29 & 59.8 & 4.9 
   & 3.09$\times$1.47, 15$\fdg$0  & 120 \\
BI37C \dotfill & 2008-09-29 & 42.5 & 2.7  
   & 1.99$\times$1.12, 25$\fdg$9 & 94 \\
BI37D \dotfill & 2008-11-28 & 43.3 & 1.0  
   & 2.31$\times$1.33, 22$\fdg$5 & 134 \\
BI37E \dotfill & 2009-02-13 & 51.6 & 16.9 
   & 3.28$\times$1.48, 16$\fdg$2  & 41 \\
BI37F \dotfill & 2009-05-19 & 61.3 & 16.9 
   & 3.28$\times$1.48, 16$\fdg$2  & 97 \\
\hline
\end{tabular}

\tablenotetext{a}{The LSR velocity at the phase-referenced spectral channel 
                  in units of \kms.} 
\tablenotetext{b}{rms noise in units of mJy beam$^{-1}$ in the emission-free 
                  spectral channel image.} 
\tablenotetext{c}{Synthesized beam size resulting from natural weighted 
                  visibilities; major and minor axis lengths and position angle.}
\tablenotetext{d}{Number of the detected maser features.}
\end{table}

\begin{deluxetable}{cr@{ $\pm$ }rr@{ $\pm$ }rr}
\tablewidth{0pt}
\tablecaption
{Parameters of the brightest \water\ maser spot from each detected maser 
feature in epoch~A to F (see Section~\ref{sec:res} for details). 
\label{tab:spots}
}
\tabletypesize{\footnotesize}
\tablecolumns{6}
\tablehead{
   \multicolumn{1}{c}{$V_{\mbox{\tiny LSR}}$\tablenotemark{a}} &
   \multicolumn{2}{c}{R.A. offset\tablenotemark{b}} & 
   \multicolumn{2}{c}{Decl. offset\tablenotemark{b}} & 
   \multicolumn{1}{r}{$I$\tablenotemark{c}}   \\ 
   \multicolumn{1}{c}{(km~s$^{-1}$)} & 
   \multicolumn{2}{c}{(mas)} & 
   \multicolumn{2}{c}{(mas)} & 
   }
\startdata
\multicolumn{6}{c}{2008 April 21 (epoch A)} \\ \hline
$-50.75$ & $-39.45$ & $0.04$ & $49.40$ & $0.09$ & $0.42$ \\
$-51.17$ & $-43.79$ & $0.07$ & $68.31$ & $0.08$ & $1.03$ \\
$-49.69$ & $-39.96$ & $0.04$ & $57.82$ & $0.04$ & $0.39$ \\
$-44.42$ & $-38.15$ & $0.04$ & $49.98$ & $0.05$ & $2.10$ \\
$-45.47$ & $-35.04$ & $0.01$ & $53.01$ & $0.04$ & $1.98$ \\
$-42.94$ & $-38.95$ & $0.01$ & $49.44$ & $0.06$ & $1.78$ \\
$-39.57$ & $-40.84$ & $0.02$ & $51.32$ & $0.07$ & $1.18$ \\
\enddata

\tablenotetext{a}{The local-standard-of-rest velocity at the intensity peak.}
\tablenotetext{b}{Position offset with respect to a selected maser feature 
   which appears in all the six epochs. 
   This reference maser feature (represented by its brightest maser spot) 
   takes the coordinates $(0,0)$ in each epoch of
   this table. The position errors from the 2D~Gaussian fitting are given as 
   well.}
\tablenotetext{c}{Peak intensity of \water\ emission of the maser feature 
   in units of Jy~beam$^{-1}$.}
\end{deluxetable}

\begin{deluxetable}{cccc}
\tablewidth{0pt}
\tabletypesize{\footnotesize}
\tablecaption{Coordinates of 54 \water\ maser features that appear 
              in any 3 consecutive epochs, selected from 
              Table~\ref{tab:spots}. They are used for the 
              proper motion calculation of individual maser 
              feature.\label{tab:proper}}
\tablehead{
   \colhead{R.A. offset\tablenotemark{a}} &
   \colhead{Dec. offset\tablenotemark{a}} & 
   \colhead{Index\tablenotemark{b}} &
   \colhead{Epoch\tablenotemark{c}} \\
   
   \colhead{(mas)} &
   \colhead{(mas)}
}
\startdata
$-38.954$&$49.436$&$1$&$A$\\
$-39.207$&$50.884$&$2$&$B$\\
$-39.857$&$53.203$&$3$&$C$\\
\hline
$-41.919$&$47.795$&$1$&$A$\\
$-42.312$&$48.766$&$2$&$B$\\
$-43.388$&$51.23$&$3$&$C$\\
\hline
$-44.362$&$33.861$&$1$&$A$\\
$-44.748$&$34.352$&$2$&$B$\\
\enddata

\tablenotetext{a}{Position offset with respect to a selected maser feature 
   which appears in all the six epochs. 
   This reference maser feature (represented by its brightest maser spot) 
   takes the coordinates $(0,0)$ in each epoch.}
\tablenotetext{b}{The maser feature identified in the earliest epoch within
                  their own group is labelled as index~1, and index~3 is
                  for the latest epoch.}
\tablenotetext{c}{The epoch in which the maser feature is found.}
\end{deluxetable}

\begin{deluxetable}{cccccc}
\tablewidth{0pt}
\tabletypesize{\footnotesize}
\tablecaption{Velocity information of the 54 \water\ maser features 
              where their proper motion can be identified according to 
              Table~\ref{tab:proper}.\label{tab:prop_vel}}
\tablehead{
   \colhead{R.A. offset\tablenotemark{a}} &
   \colhead{Dec. offset\tablenotemark{a}} &
   \colhead{$V_{\mathrm{R.A.}}$\tablenotemark{b}} &
   \colhead{$V_{\mathrm{Dec.}}$\tablenotemark{c}} & 
   \colhead{$V_{\mathrm{LSR}}$\tablenotemark{d}} &   
   \colhead{$V_{\mathrm{3D}}$\tablenotemark{e}} \\

   \colhead{(mas)} &
   \colhead{(mas)} &
   \colhead{(\kms)} &
   \colhead{(\kms)} &
   \colhead{(\kms)} &
   \colhead{(\kms)}
}
\startdata
\multicolumn{6}{c}{Jet~1} \\ \hline
$-38.95$&$49.44$&$-41.70$&$173.95$&$-42.94$&$183.96$\\
$-41.92$&$47.80$&$-67.83$&$158.62$&$-37.03$&$176.44$\\
$-44.36$&$33.86$&$-53.24$&$61.78$&$-13.61$&$82.69$\\
$-38.58$&$94.80$&$-47.79$&$167.85$&$-0.53$&$174.53$\\
$-38.58$&$94.80$&$-39.80$&$159.40$&$-0.53$&$164.30$\\
$-34.09$&$-6.06$&$-87.60$&$106.00$&$56.65$&$148.72$\\
$-29.31$&$91.16$&$-21.24$&$55.32$&$1.58$&$59.28$\\
\enddata

\tablenotetext{a}{Position offset of the maser features with index~1 as shown 
   in Table~\ref{tab:proper}, with respect to a selected maser feature 
   which appears in all the six epochs. 
   This reference maser feature (represented by its brightest maser spot) 
   takes the coordinates $(0,0)$ in each epoch.}
\tablenotetext{b}{The proper motion velocity along R.A. direction.}
\tablenotetext{c}{The proper motion velocity along Dec. direction.}
\tablenotetext{d}{The local-standard-of-rest velocity at the intensity peak.}
\tablenotetext{e}{The estimated 3-dimensional velocity.}

\end{deluxetable}

\begin{deluxetable}{lcccccccc}
\tablewidth{0pt}
\tabletypesize{\footnotesize}
\tablecaption
{Parameters of the precessing ballistic jets in different models.
The errors give the 68\% confidence level for each parameter.
The outflow velocity of jet~1 is taken to be $138$\kms, 
while that of jet~2 is $99$\kms. 
These values are estimated from the proper
motion of \water\ maser features.\label{tab:model}}
\tablehead{
   \colhead{Jet} &
   \colhead{$V_{\mathrm{sys}}$\tablenotemark{a}} &
   \colhead{$T$\tablenotemark{b}} & 
   \colhead{$\alpha$\tablenotemark{c}} &   
   \colhead{Age\tablenotemark{d}} &
   \colhead{P.A.\tablenotemark{e}} &   
   \colhead{I.A.\tablenotemark{f}} &          
   \colhead{$R^{2}$} &
   \colhead{Note\tablenotemark{g}} \\
    
   & 
   \colhead{(\kms)} &
   \colhead{(years)} &
   \colhead{(degrees)} & 
   \colhead{(years)} &
   \colhead{(degrees)} &   
   \colhead{(degrees)}
}
\startdata
$1$ & $47.44\pm0.61$ & $56.03\pm1.28$ & $28.04\pm0.29$ & $\sim$19 & $-17.22\pm0.64$ & $-2.13\pm0.35$ & $0.86$ \\
$2a$ & $60.50\pm0.51$ & $73.32\pm3.44$ & $18.91\pm0.74$ & $\sim$30 & $11.70\pm0.73$ & $21.69\pm1.26$ & $0.73$ & s, r\\
$2b$ & $54.28\pm0.60$ & $52.57\pm1.25$ & $22.17\pm0.58$ & $\sim$30 & $17.07\pm0.66$ & $19.15\pm0.93$ & $0.54$ & s, u\\
$2c$ & $55.88\pm0.44$ & $94.90\pm2.21$ & $17.97\pm0.38$ & $\sim$26 & $-7.80\pm0.55$ & $-4.39\pm1.83$ & $0.70$ & o, r\\
\enddata

\tablenotetext{a}{The systemic velocity of the driving source.}
\tablenotetext{b}{Precession period of the jet.}
\tablenotetext{c}{Precession angle of the jet.}
\tablenotetext{d}{Dynamical age of the jet. It is calculated from the 
                  distance between the tip and the source of the jet, divided
                  by the jet velocity.}
\tablenotetext{e}{Position angle from north to east of the jet axis.}
\tablenotetext{f}{Inclination angle between the jet axis and the sky plane.
                  Positive value means the northern end of the 
                  axis is pointing out from the sky plane, and negative means
                  the southern end of the axis is pointing out from the 
                  sky plane.}
\tablenotetext{g}{Note ``s'' means in this model jet~2 is having the same 
                  counter-clockwise precessing
                  direction as jet~1, while ``o'' means opposite precessing
                  direction. Note ``r'' means the jets have two different 
                  (resolvable) driving sources, while ``u'' means the driving 
                  sources are unresolvable and just appear to be at the same 
                  point.}
\end{deluxetable}

\end{document}